\documentclass[longauth]{aa}
\usepackage{natbib,twoopt}
\usepackage{graphicx}
\usepackage{marvosym}
\bibpunct{(}{)}{;}{a}{}{,} 
\newcommand{\degree}{$^{\circ}$}

\newcommand{\micron}{$\mathrm{\mu m}$}

\title{High-Contrast study of the candidate planets and protoplanetary disk around HD~100546\thanks{Based on data collected at the European Southern Observatory, Chile (ESO Programs 095.C-0298, 096.C-0241, 096.C-0248, 097.C-0523, 097.C-0865 and 098.C-0209).}}
\author{E. Sissa\inst{1,2},
R. Gratton\inst{1}, 
A. Garufi\inst{3,4}, 
E. Rigliaco\inst{1},  
A. Zurlo\inst{5,6} ,
 D. Mesa\inst{1,7}, 
 M. Langlois\inst{7,8}, 
 J. de Boer\inst{9}, 
 S. Desidera\inst{1},  
 C. Ginski\inst{9}, 
 A.-M. Lagrange\inst{10}, 
 A.-L. Maire\inst{11}, 
 A. Vigan\inst{6}, 
 M. Dima\inst{1}, 
 J. Antichi\inst{1,12},
 A. Baruffolo\inst{1},
 A. Bazzon\inst{3}, 
 M. Benisty\inst{10},
 J.-L. Beuzit\inst{10}, 
 B. Biller\inst{11,13}, 
 A. Boccaletti\inst{14}, 
 M. Bonavita\inst{1,13}, 
 M. Bonnefoy\inst{10}, 
 W. Brandner\inst{11},
 P. Bruno\inst{15},  
 E. Buenzli\inst{3},
 E. Cascone\inst{16},
 G. Chauvin\inst{10,17}, 
 A. Cheetham\inst{18}, 
 R.U. Claudi\inst{1},
 M. Cudel\inst{10}, 
 V. De Caprio\inst{16},
 C. Dominik\inst{9}, 
 D. Fantinel\inst{1},
 G. Farisato\inst{1},
 M. Feldt\inst{11},
 C. Fontanive\inst{13}, 
 R. Galicher\inst{14}, 
 E. Giro\inst{1,19},
 J. Hagelberg\inst{10}, 
 S. Incorvaia\inst{20},
 M. Janson\inst{11,21}, 
 M. Kasper\inst{10,22}, 
 M. Keppler\inst{11}, 
 T. Kopytova\inst{11}, 
 E. Lagadec\inst{23}, 
 J. Lannier\inst{10}, 
 C. Lazzoni\inst{1,2},
 H. LeCoroller\inst{6}, 
 L. Lessio\inst{1},
 R. Ligi\inst{6}, 
 F. Marzari\inst{1},
 F. Menard\inst{10}, 
 M.R. Meyer\inst{3,24}, 
 D. Mouillet\inst{10}, 
 S. Peretti\inst{18}, 
 C. Perrot\inst{14}, 
 P. J. Potiron\inst{23}, 
 D. Rouan\inst{14},
 B. Salasnich\inst{1}, 
 G. Salter\inst{6},
 M. Samland\inst{11}, 
 T. Schmidt\inst{14},
 S. Scuderi\inst{15},
 F. Wildi\inst{18}}
\authorrunning{E. Sissa et al.}
\offprints{E. Sissa,  \\
   \email{elena.sissa@oapd.inaf.it} }
\institute{
 INAF-Osservatorio Astronomico di Padova,  Vicolo dell'Osservatorio 5, I-35122, Padova, Italy 
\and Dipartimento di Fisica e Astronomia - Universita' di Padova, Vicolo dell'Osservatorio 3, I-35122, Padova, Italy  
\and Institute for Particle Physics and Astrophysics, ETH Zurich, Wolfgang-Pauli-Strasse 27, CH-8093 Zurich, Switzerland 
\and Universidad Auton\'{o}ma de Madrid, Dpto. F\'{i}sica Te\'{o}rica, M\'{o}dulo 15, Facultad de Ciencias, Campus de Cantoblanco, 28049, Madrid, Spain 
\and N\'ucleo de Astronom\'ia, Facultad de Ingenier\'ia, Universidad Diego Portales, Av. Ejercito 441, Santiago, Chile  
\and Aix-Marseille Universit\'e, CNRS, LAM (Laboratoire d'Astrophysique de Marseille) UMR 7326, 13388, Marseille, France 
\and University of Atacama, Copayapu 485, Copiapo, Chile 
\and CRAL, UMR 5574, CNRS, Universit\'e de Lyon, Ecole Normale Suprieure de Lyon, 46 Alle d'Italie, F-69364 Lyon Cedex 07, France 
\and Leiden Observatory, Leiden University, PO Box 9513, 2300 RA Leiden, The Netherlands 
\and Universit\'{e} Grenoble Alpes, CNRS, IPAG, 38000 Grenoble, France 
\and Max-Planck-Institut f\"{u}r Astronomie, K\"{o}nigstuhl 17, D-69117 Heidelberg, Germany 
\and INAF-Osservatorio Astrofisico di Arcetri,  L.go E. Fermi 5, 50125 Firenze, Italy 
\and Institute for Astronomy, University of Edinburgh, Blackford Hill, Edinburgh EH9 3HJ, UK 
\and LESIA, Observatoire de Paris-Meudon, CNRS, Universit\'{e} Pierre et Marie Curie, Universit\'{e} Paris Diderot, 5 Place Jules Janssen, F-92195 Meudon, France 
\and INAF-Osservatorio Astrofisico di Catania, Via S. Sofia 78, I-95123 Catania, Italy 
\and INAF-Osservatorio Astronomico di Capodimonte, Via Moiariello,16 I-80131 Napoli, Italy 
\and Unidad Mixta Internacional Franco-Chilena de Astronomia,
CNRS/INSU UMI 3386 and Departamento de Astronomia,, Universidad
de Chile, Casilla 36-D, Santiago, Chile 
\and Observatoire de Gen\'{e}ve, University of Geneva, 51 Chemin des Maillettes, 1290, Versoix, Switzerland 
\and INAF-Osservatorio Astronomico di Brera, via Emilio Bianchi 46, 23807, Merate (LC), Italy 
\and INAF-Istituto di Astrofisica Spaziale e Fisica Cosmica di Milano, Via E. Bassini 15, 20133 Milano, Italy 
\and Department of Astronomy, Stockholm University, SE-106 91 Stockholm, Sweden 
\and European Southern Observatory, Karl-Schwarzschild-Str. 2, D85748 Garching, Germany  
\and Laboratoire Lagrange (UMR 7293), UNSA, CNRS, Observatoire de  la  C\^{o}te  d'Azur, Bd. de l'Observatoire, 06304 Nice Cedex 4, France 
\and Department of Astronomy, University of Michigan, 311 West
Hall, 1085 S. University Avenue, Ann Arbor, MI 48109, USA 
}

\date{Received  /
Accepted }

\abstract{The nearby Herbig Be star HD~100546 is known to be a laboratory for the study of protoplanets and their relation with the circumstellar disk, that is carved by at least two gaps. We observed the HD~100546 environment with high contrast imaging exploiting several different observing modes of SPHERE, including data sets with/without coronagraphs, dual band imaging, integral field spectroscopy and polarimetry. The picture emerging from these different data sets is complex. Flux-conservative algorithms images clearly show the disk up to 200~au. More aggressive algorithms reveal several rings and warped arms that are seen overlapping the main disk. The bright parts of this ring are found out to lie at considerable height over the disk mid-plane at about 30~au. Our images demonstrate that the brightest wings close to the star in the near side of the disk are a unique structure, corresponding  to the outer edge of the intermediate disk at $\sim40$~au.  Modelling of the scattered light from the disk with a geometrical algorithm reveals that a moderately thin structure ($H/r$=$0.18$ at 40 au)  can well reproduce the light distribution in the flux-conservative images. We suggest that the gap between 44~au and 113~au span between the 1:2 and 3:2 resonance orbits of a massive body located at $\sim 70$~au, that might coincide with the candidate planet HD~100546b detected with previous thermal IR observations. In this picture, the two wings can be the near side of a ring formed by disk material brought out of the disk at the 1:2 resonance with the same massive object. While we find no clear evidence confirming detection of  the planet candidate HD~100546c in our data, we find a diffuse emission close to the expected position of HD~100546b. This source can be described as an extremely reddened substellar object surrounded by a dust cloud or its circumplanetary disk. Its astrometry is broadly consistent with a circular orbital motion on the disk plane, a result that could be confirmed with new observations. Further observations at various wavelengths are required to fully understand the complex phenomenology of HD~100546. 
}

\begin{document}
\maketitle

\section{Introduction}
The number of known exoplanets around main-sequence stars is increasing rapidly. However, to date, the number of detected (forming) exoplanets around pre-main-sequence stars is still low. Possibly the best examples are LkCa 15 \citep{kraus2012, sallum2015Nat}, HD~169142 \citep{biller2014, reggiani2014}, HD~100546 \citep{2013Quanz1, 2014Currie, 2015Quanz, 2015Currie, rameau2017} and PDS 70 \citep{keppler2018,mueller2018}. To understand the formation process of planets, we need to study the initial condition and evolution of circumstellar disks, and how the disk can be shaped by ongoing planet formation.

Recently developed high-contrast and high-angular resolution imaging instruments such as SPHERE \citep[Spectro-Polarimetric High-contrast Exoplanet REsearch;][]{beuzit2008}, GPI \citep[Gemini Planet Imager;][]{macintosh2014}, SCExAO \citep[Subaru Coronagraphic Extreme Adaptive Optics, ][]{SCExAO} and FLAO \citep[First Light AO system,][]{FLAO} provide the excellent capability to directly obtain images of protoplanetary disks around nearby young stars, up to the inner tens of au, in scattered light and thermal emission. This makes sometime possible to observe planets in their birthplace investigating their interaction with the disk. Imaging protoplanetary disks around young stars allows us to detect spiral structures and gaps which might be produced  by gravitational perturbation of forming planets, as demonstrated by  e.g. \cite{grady2001,thalmann2010,garufi2013,2015Pinilla,dong2016}.

HD~100546 is a nearby \citep[$d$=$ 109 \pm 4$ pc,][]{gaia2016a} well studied Herbig Be star with spectral type  B9Vne \citep{2006MNRAS.371..252L} which harbours a large disk. Several studies have been conducted in the past years on this disk, from spectroscopic and photometric analysis to direct imaging. The first evidence of the presence of the disk around HD~100546 was obtained by \cite{1989Hu} measuring the IR excess in the  SED. The mid-IR excess in the SED of this source requires a thickening of the disk at $\sim0.1\arcsec$, that can be explained by a proto-Jupiter carving a gap \citep{bouwman2003}. This idea is also supported by far-ultraviolet long-slit spectroscopy with HST/STIS, that detected a central cavity up to 0.13\arcsec\ \citep{grady2005}, by UVES observations of [OI] emission region \citep{acke2006}, using spectro-astrometry with CO roto-vibrational emission by \cite{vanderplas2009} and by AMBER/VLTI observations in the K-band \citep{benisty2010}.

The disk was first imaged in scattered light in the J and K bands with the adaptive optics system ADONIS coupled with a pre-focal optics coronagraph \citep{2000Pantin}. The disk was detected up to 2\arcsec\ from the star, with a density peak at $\sim40$\,au. Many successive works revealed the complexity of its geometry by means of both scattered light images \citep[e.g.,][]{augereau2001, 2007Ardila, 2011Quanz, 2016Garufi, follette2017} and also, (sub-)millimeter images \citep[e.g.,][]{walsh2014, 2014Pineda, 2015Wright}. The radial profile of the disk brightness becomes less steep at separations $<2.7$\arcsec\ in both HST/NICMOS2 H-band \citep{augereau2001} and  ADONIS $Ks$ \citep{grady2001} band. However this change is not visible in the optical. Moreover, the semi-minor axis brightness profile is asymmetric in the H-band  \citep{augereau2001}; this can be related to an optically thick circumstellar disk inside 0.8\arcsec\ and an optically thin disk at larger separations.

Thanks to the higher angular resolution reachable with HST/STIS coronagraphic images, the first disk structures were detected: spiral dark filaments appear mostly in the SW at separations greater then 2.3\arcsec \citep{grady2001}. Following observations revealed additional spiral structures both at short and at far separations, possibly related to forming planets \citep{2007Ardila,boccaletti2013,2015Currie, 2016Garufi}. The first Polarimetric Differential Imaging (PDI) detection of the disk \citep{2011Quanz} in H and $\mathrm{K_s}$ NACO filters resolved the disk between 0.1\arcsec\ and 1.4\arcsec, locating the disk inner rim at 0.15\arcsec. They noted an asymmetry along the brightness profile that they interpreted as the presence of two dust populations. Deeper observations by \cite{avenhaus2014} reached the innermost region ($\sim$ 0.03\arcsec) and found a spiral arm in the far-side of the disk. The disk is found out to be strongly flared in the outer regions  ($>$ 80-100 au) and it is expected to generate shadowing effect on the forward-scattering side of the disk. The disk appears faint and red, due to the combination of particles sizes and disk geometry as demonstrated by \cite{2013Mulders} and later confirmed by \cite{stolker2016} comparing models to PDI data.

In addition to these data from the disk, several studies suggest that HD~100456 may host at least two planets \citep{2013Quanz1,2014Brittain,2015Quanz,2015Currie} located at $\sim 55$\,au (hereafter CCb) and $\sim13$\,au (hereafter CCc)  from the central star. This makes the system a powerful laboratory to study planet formation and planet-disk interaction.

\citet{2015Quanz} inferred the temperature and emitting radius of CCb in the $L'$ and $M$ bands and concluded that what they imaged could be a warm circumplanetary disk rather than the planet itself. An extended emission at near-IR wavelengths, possibly linked with this candidate planet was also detected by \citet{2015Currie} centred at  $0.469\pm0.012$\arcsec, $PA$=$7.04$\degree $\pm 1.39$\degree. Its infrared colors are extremely red compared to models, indicating that it could still be embedded by accreting material.

The existence of CCc was proposed by \citet{2013Brittain}, who ascribed the variability of the CO ro-vibrational lines to a compact source of emission at $\sim$15 au from the central star. \cite{2015Currie} showed that this object is detected in GPI H-band images. However there is a debate on this result, because detection strongly depends on the reduction method used  \citep[see e.g.][]{2016Garufi,follette2017,currie2017} and on its nature, since it can be interpreted with alternative equally plausible origins like disk hot spot or indirect signature of the presence of the planet.

First observations of HD~100546 with VLT/SPHERE were presented in \cite{2016Garufi}, where it was found that the disk around HD 100546 is truncated at about 11 au and the cavity is consistent with being intrinsically circular. More recent works by \cite{follette2017,rameau2017, currie2017}, all based on GPI data sets, debate on the presence of both the planets and the impact of the reduction methods on them.

HD~100546 is located in the Sco-Cen complex. In this analysis we consider a stellar mass of $M_* $=$2.4 M_\odot$ \citep{2014Brittain}, and apparent magnitudes of $J$=$6.42$, $H$=$5.96$, and $K$=$5.42$ \citep{2MASS}, as well as $L'$=$4.52$ and $M'$=$4.13$ as in \citet{2015Quanz}. For the disk, we assume an inclination $i$=$42$\degree\ \citep{2007Ardila,2014Pineda} and $PA$=$146$\degree\ \citep{2014Pineda}.

In this paper, we present the results of a two year observational campaign on this star, based on observations in direct imaging with VLT/SPHERE, the new high-contrast and high-resolution instrument dedicated to exoplanet and disk imaging, that includes both pupil tracking and polarimetric observations. SPHERE includes a powerful extreme adaptive optics system \citep[Sphere AO for eXoplanet Observation, SAXO,][]{fusco2006}, providing a currently unmatched Strehl Ratio of up to 92\% in the H-band for bright (R < 9) sources,  various coronagraphs \citep[see][]{martinez2009,carbillet2011}, an infrared differential imaging camera \citep[IRDIS,][]{dohlen2008}, an infrared integral field spectrograph \citep[IFS,][]{claudi2008} and a visible differential polarimeter and imager \citep[ZIMPOL,][]{thalmann2008}.

The paper is structured as follows. The observing strategy, sky conditions and the reduction of these data set are described in Sect. \ref{sec:observations}. The results for the disk and planets are given in Sect. \ref{sec:disk} and \ref{sec:planets} respectively. Conclusions are given in Sec.~\ref{sec:concl}.

\section{Observations and data reduction}
\label{sec:observations}

\subsection{Observations}
\begin{table*}
\begin{minipage}{\textwidth}
\centering
\caption[]{Observations summary}
 \begin{tabular}{lccccccccc}
\hline
Date & Obs. mode & $t_{exp}$ [s]& {Rot [\degr]} & {$r_C$ [mas]}& {ND Filter} & $\sigma$ [\arcsec] & $\tau_0$ [ms] & SR & $ 5\sigma$ \MVAt  $0.5$\arcsec  \\
\hline
\hline
  May  3 2015 & IRDIFS\_EXT & 3840 & 22.32 & 92.5 & no  & 
 0.75 &2.1 &  0.65& 13.21 \\
  May 29 2015 & IRDIFS      & 6048 & 36.37 & 92.5 & no  & 0.70 &{1.9}& 0.80 & 13.88\\
  Jan 17 2016 & IRDIFS\_EXT & 4096 & 22.23 & 92.5 & no  & 1.54 & 1.7 &0.80 & 12.88\\
  Mar 26 2016 & IRDIFS\_EXT & 4096 & 22.25 & 92.5 & no  & 2.58 &0.9 &  0.57 & 12.68\\
  Mar 31 2016 & IRDIS DPI J & 1280 & - & 92.5 & no & 1.98 & 0.93& 0.85 & \\
  Apr 16 2016 & IRDIFS\_EXT & 4512 & 29.42 & no   & 2.0 & 0.66 & 3.3& 0.78& 10.11\\ 
  May 25 2016 & IRDIS DPI K & 2304 & - & 92.5 & no & 0.75 &  5.10 & na & \\
  May 31 2016 & IRDIFS\_EXT & 4400 & 28.95 & 92.5 & no  & 0.75 & 2.4 & 0.55& 13.35\\
  Feb  7 2017 & IRDIFS\_EXT & 5280 & 28.38& no & 2.0 & 0.78 & 6.7 & 0.84 & 10.55\\
    \hline
    \end{tabular}%
\label{tab:data}
\end{minipage}
\tablefoot{ Date, SPHERE observing mode,  Total integration time ($t_{exp}$), Total field rotation (Rot), Coronagraph radius ($r_C$), Neutral Density Filter, average DIMM seeing FWHM on source in V band ($\sigma$), the average coherence time $\tau_0$, the  average Strehl ratio in H band (SR) coming from SPARTA data are presented. The deepest $5\sigma$ contrast limit reached at 0.5\arcsec with IFS after spectral ADI is also given to show the relative quality of the data sets.}
\end{table*}

We observed HD~100546 at different epochs as part of the SHINE (SpHere INfrared survey for Exoplanets) Guaranteed Time Observations (GTO) program using IRDIS and IFS simultaneously: the IRDIFS mode (with IFS operating between 0.95 and 1.35 \micron\ and IRDIS working in in dual imaging mode in the H2H3 filter pair at 1.59 and 1.67 \micron, \citealt{vigan2010} ) and the IRDIFS\_EXT mode (with IFS operating at Y-H wavelengths 0.95-1.65 \micron\ and IRDIS in the K1K2 band filters at 2.11 and 2.25 \micron ). These observations were all carried-out with the \texttt{N\_ALC\_YJH\_S} apodized Lyot coronagraph (inner working angle, IWA $\sim 0.1$\arcsec, \citealt{boccaletti2008}) except for two sequences which were acquired without coronagraph in order to study the central region (at less than 0.1\arcsec\ from the star). In this case, to avoid heavy saturation of the star, a neutral density filter with average transmission $\sim$1/100 was used. All data were taken in pupil stabilized mode in order to perform Angular Differential Imaging \citep[ADI, see e.g.\,][]{2006Marois} for subtracting the stellar residuals.

HD~100546 was also observed in Polarimetric Differential Imaging \citep[PDI, see e.g.\,][]{kuhn2001, 2011Quanz}, as part of the DISK GTO program, using IRDIS with the same coronagraphic mask as for the classical imaging. The source was observed for a total time of 21 minutes in the J band ($1.26$ \micron) and 38 minutes in the K band ($2.181$ \micron ) during the night of March 31st, 2016, and May 25th, 2016, respectively. The orientation of the derotator was chosen in order to optimize the polarimetric efficiency of IRDIS (de Boer et al., in prep.).

Several of the observations were carried-out under moderate and good weather conditions (with an average coherence time, $\tau_0$, longer than 2.0 milliseconds). Details on the full set of observations are presented in Table \ref{tab:data},  $\tau_0$ and Strehl ratio were measured by SPARTA, the ESO standard real-time computer platform that controls the AO loop.
\begin{figure}
\centering
\includegraphics[width=\columnwidth]{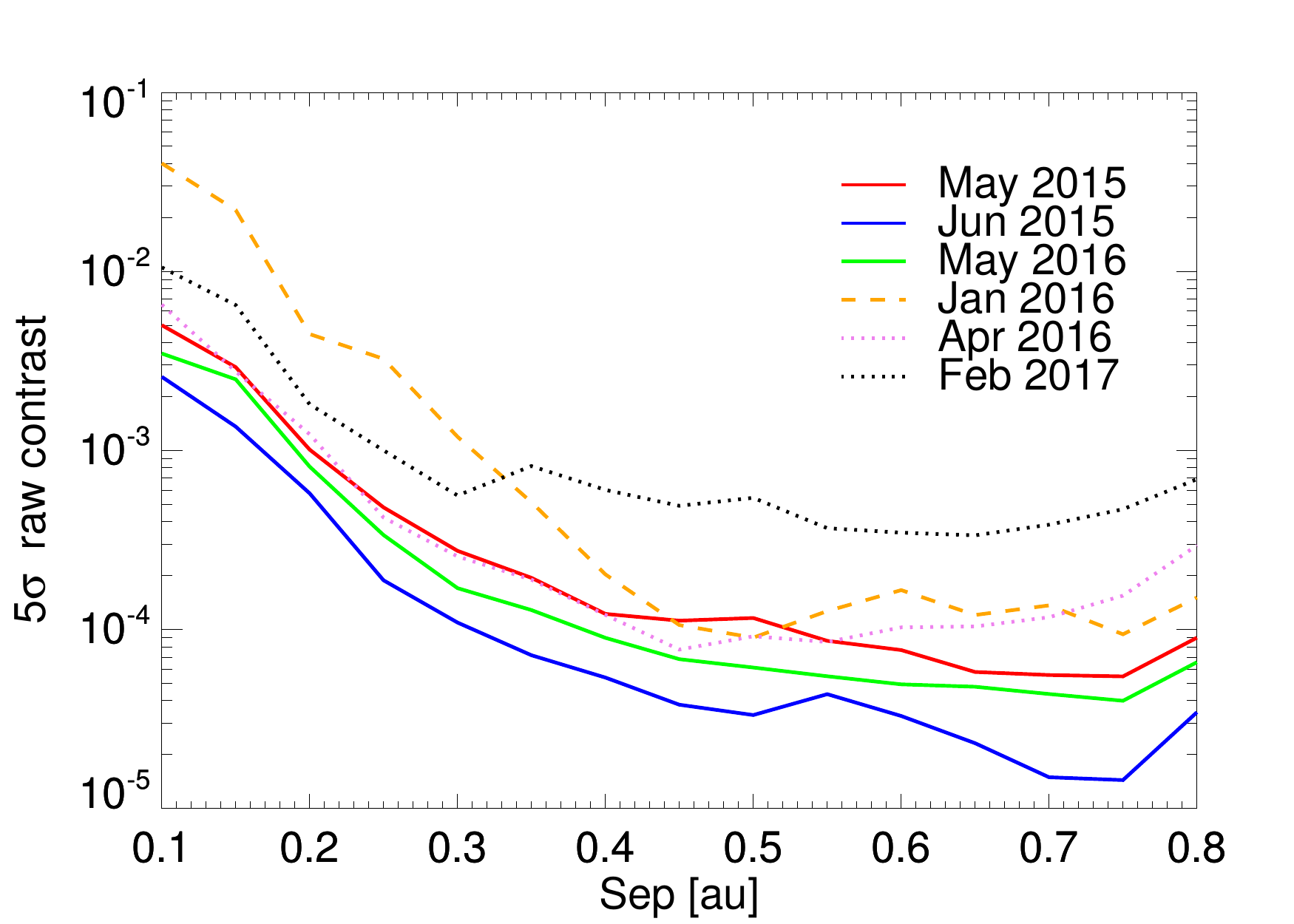}
\caption{$5\sigma$ raw contrast of the IFS images at all epochs but March 2016. }
\label{raw_contrast}
\end{figure}

In order to access the quality of the observational material, we plotted in Fig. \ref{raw_contrast} the contrast we achieved in the IFS raw images as a function of the separation for all the non polarimetric images but March 2016 data set, due to its very unstable sky conditions. The contrast plotted here were obtained combining the root mean square scatter of a 1$\lambda/D$ diameter spot along a 1$\lambda/D$ wide annulus, at different separation from the star as described in \cite{mesa2015}. These contrast values were then corrected for the low number statistics according to \cite{mawet2014}. We noticed that these raw image contrast are comparable to those obtained with the two GPI data sets presented in \cite{currie2017}.

We found that the highest quality coronagraphic observations are those from May 2015 and May 2016 while the observations obtained in January and March 2016 are of poor quality due to the atmospheric conditions or low system performances (calibration of the deformable mirror voltages was not optimal in January 2016). For the non-coronagraphic data sets, the lack of a coronagraph implies much stronger diffraction patterns and the need of using the neutral density filter, common to IFS and IRDIS, to avoid saturation close to the center. This strongly reduces the signal; for this reason the sensitivity of these images is far from optimal at separations larger than 0.1\arcsec\ but they allow access to the closest separations with unprecedented spatial resolution.

\subsection{IRDIFS and IRDIFS\_EXT data reduction}
\label{sec:ADIobs}

We performed the basic data reduction of IRDIS and IFS (bad pixel removal, flat fielding, image alignment, sky subtraction) with version 0.15.0 of SPHERE Data Reduction and Handling (DRH) pipeline \citep{pavlov2008}. Further elaboration of the images (deconvolution for lenslet-to-lenslet cross talk, refinement of the wavelength calibration, correction for distortion, fine centering of the images, frame selection) was performed at the SPHERE Data Center (DC) in Grenoble\footnote{\url{http://sphere.osug.fr/spip.php?rubrique16&lang=en}} \citep{delorme2017}. Additional details on the adopted procedures are described in \cite{zurlo2014}, \cite{mesa2015} and \cite{maire2016astrom}. We then applied various algorithms for differential imaging such as classical ADI \citep[cADI,][]{ 2006Marois}, template-LOCI \citep[TLOCI,][]{2014Marois}, and Principal Component Analysis \citep[PCA,][]{soummer2012, amara2012}. These procedures are available at the SPHERE DC through the SpeCal software \citep{galicher2018}. 
 
For IFS, the cADI is based on a median combination of the images, while the PCA is evaluated on the whole image at the same time; no exclusion zones are considered. 

For IRDIS, no spatial filtering was applied to the data beforehand. We performed several reduction and, verified that both disk and wide companions (see Appendix~\ref{sec:bkg_objects}) are detected using the alternate methods. We also confirm that these detections are robust across a range of algorithm parameter space (e.g. varying numbers of principal components from 2 to 5). In classical ADI, SpeCal averages the cube of frames over the angular dimension for each spectral channel to remove the speckle contamination. After a rotation is applied, the frames are averaged using mean or median combination to produce one final image per spectral channel. The median-combination is applied for disk images because it is less sensitive to uncorrected hot/bad pixels.  When estimating the candidate companions photometry the mean is used instead of the median to preserve linearity. PCA is evaluated on the whole image at the same time; no exclusion zones are considered and the number of PCA modes range from 2 to 5. There is no frame selection to minimize the self-subtraction of point-like sources when deriving the principal components. The principal components are calculated for each spectral channel independently. Each frame is then projected onto the first 2 to 5 components to estimate the speckle contamination. For TLOCI reduction, the minimum residual flux of a putative companion, because of self subtraction, is at least 15\% of the candidate flux. The minimum radius where the speckle are calibrated and subtracted is 1.5 FWHM using annular sections. The maximum number of frames to estimate the speckles is 80, no singular value decomposition cutoff was used. The speckle contamination is estimated by linear combination of frames  that minimizes the residual energy in the considered region using the bounded variables least-squares algorithm by Lawson \& Hanson (1995). The contrast curves calculation assumes a flat spectra for the companion candidates. 

Although Specal Software can perform spectral differential imaging \citep[SDI, ][]{racine1999}, we use procedures that treat each individual spectral channel separately. We also performed Reference Star Differential Imaging (RDI), subtracting to each datacube frame a reference PSF obtained from the star observed during the same or a close night, with the same set-up that is the most similar, in terms of luminosity and noise model, to our target. HD~95086 is ideal for our purpose since the debris disk and close companion around this star are faint \citep{chauvin2018} and do not impact on the photometry of HD~100546. The RDI was performed using as reference the observations of HD~95086 taken on May 3rd, 2015.

In addition, we used a set of customized data analysis procedures set up for performing the monochromatic PCA, taking into account the effect of the self-subtraction.

\subsection{IRDIS PDI data reduction}
\label{sec:PDIobs}
In IRDIS PDI observations, the stellar light is split into two beams with perpendicular polarization states. A half-wave plate allows to shift the orientation of the polarization four times by $22.5$\degree\ in order to obtain a full set of polarimetric Stokes vectors. The data presented in this work are reduced following the double difference method \citep{kuhn2001}  as described by \citet{ginski2016}. The resulting $Q$ and $U$ parameter are finally combined to obtain the polar Stokes vector $Q_\phi$ and $U_\phi$ as from:
\begin{equation}
Q_{\phi} = +Q \cos{(2\phi)} + U \sin{(2\phi)}
\end{equation}
\begin{equation}
U_{\phi} = -Q \sin{(2\phi)} + U \cos{(2\phi)}
\end{equation}
where $\phi$ is the position angle of the location of interest \citep{schmid2006}. With these definitions, positive $Q_\phi$ values correspond to azimuthally polarized light, while negative signal is radially polarized light. $U_\phi$ contains all signal with 45\degree\ offset from radial or azimuthal.

\section{The Disk} 
\label{sec:disk}

\begin{figure*}
\centering
\begin{tabular}{rl}
\includegraphics[trim={3.8cm 12cm 3.8cm 1cm},clip,width=0.5\textwidth]{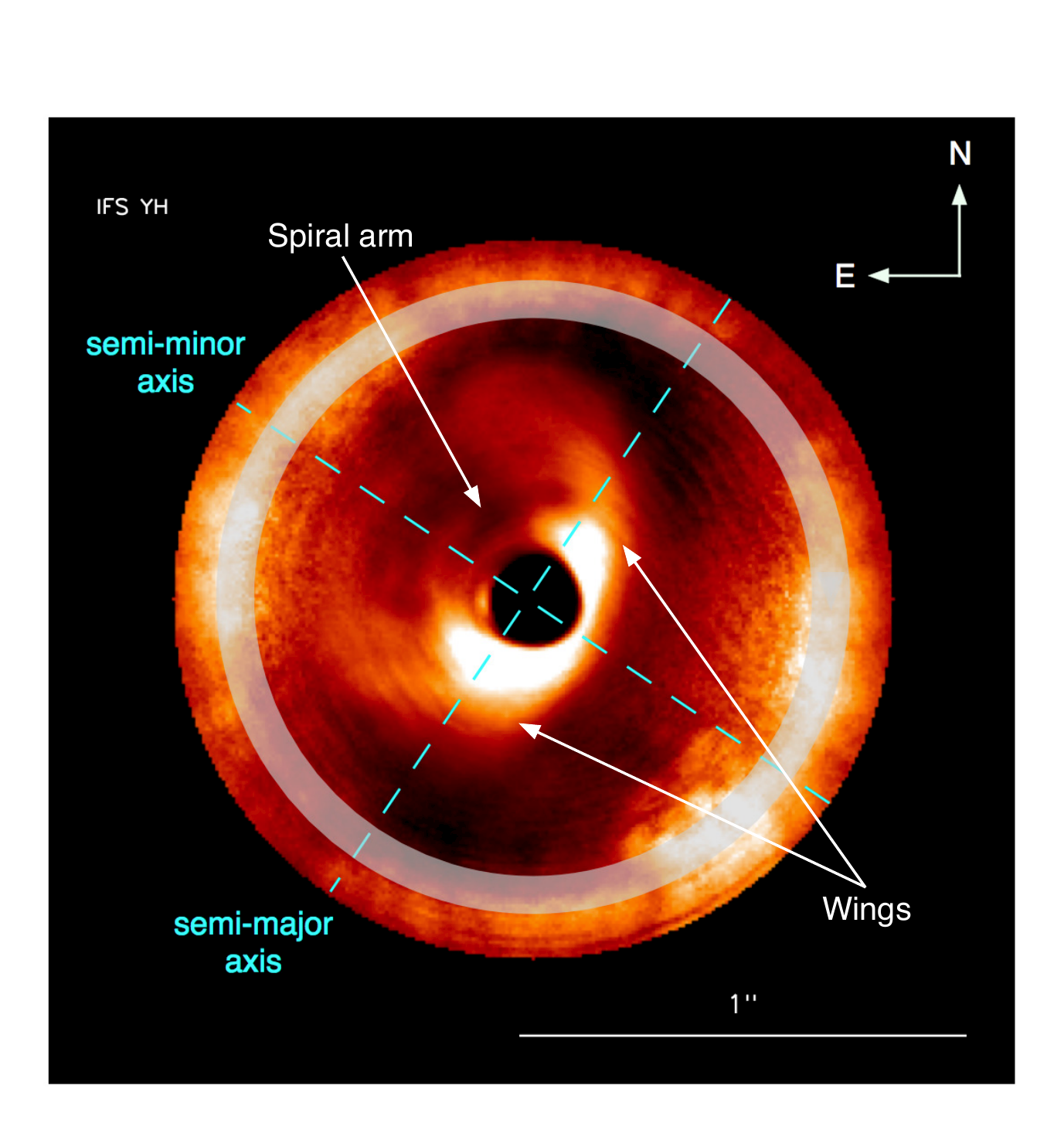}& 
\includegraphics[trim={3.8cm 12cm 3.8cm 1cm},clip,width=0.5\textwidth]{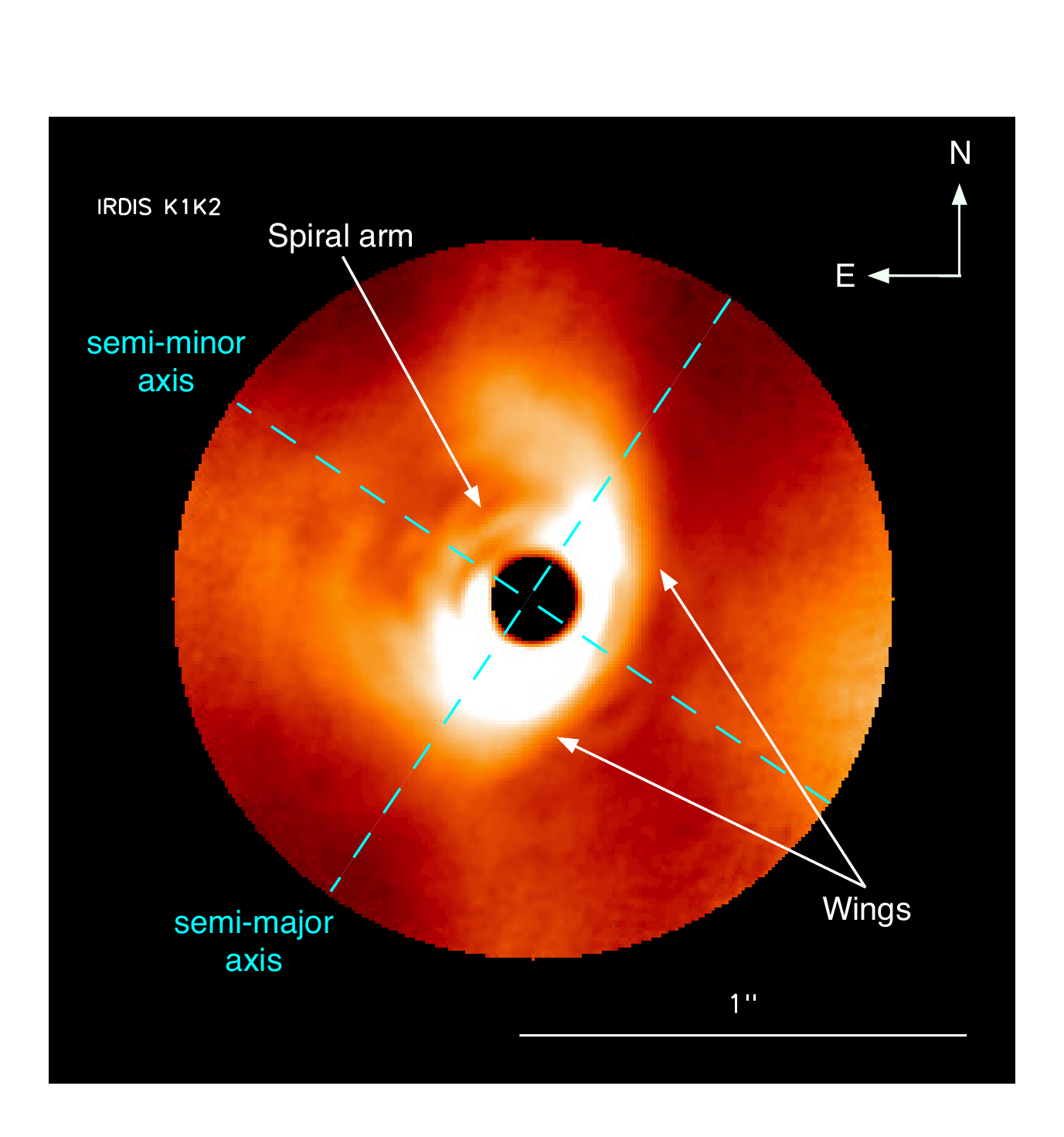}\\
\end{tabular}
\caption{May 2015  IFS collapsed YH (left) and IRDIS K1K2 (right) results for HD100546 using RDI. Both images are $r^2$-scaled to also enhance structures at larger radii. The grey circle in the left image indicate the SPHERE AO control radius at $r\approx 20 \times \lambda /D$.}
\label{IRDIS_IFS_RDI}
\end{figure*}

\subsection{Disk morphology}

\subsubsection*{Intensity Images}

Figure \ref{IRDIS_IFS_RDI} shows the central part of the HD~100546 disk as obtained applying the RDI technique both for IFS and IRDIS. The light distribution appears elliptical, oriented in agreement with PA=146\degree\ and the ratio between major and minor axis is consistent with an inclination of 42\degree\ as found by \cite{2007Ardila} using HST/ACS observations of HD~100546. The South-East (SE) part in both images appears brighter than the North-West (NW) part. The South-West (SW) part of the disk along the minor axis represents the near side to the observer and shows a strong depletion in the light distribution at a separation of $>0.2$\arcsec. The \textit{two wings} detected in the H2H3 and K1K2 bands by \cite{2016Garufi} are also distinguishable at IFS shorter wavelengths, especially the northern one. Hereafter, we will keep \cite{2016Garufi} nomenclature for these two structures since they appear symmetric with respect to the minor axis, and do not have the same concavity, as expected in the case of multiple spirals disk. In these images we do not detect the spiral arm seen by \cite{avenhaus2014}, but in the IRDIS K1K2 data there is a thin spiral arm like structure, almost circular, at separation of 0.2\arcsec, from PA$\sim100$\degree\ to PA$\sim4$\degree\ that is barely visible also in the IFS images. No relevant features are detected at separations greater then $\sim0.8$\arcsec\ in both these images, so that they are masked. The bright ring seen in IFS data at a separation of 0.65\arcsec\ corresponds to the SPHERE AO control radius\footnote{Within $r \approx 20 \times \lambda / D$ (with 20 being half the number of deformable mirror actuators along a side) the SPHERE AO system efficiently suppresses the PSF down to the level of a residual stellar halo. At this radius, the AO-corrected image may show a circular artifact which is highlighted by the spatial filtering.}. The control radius at the wavelengths of IRDIS K1K2 bands is at a separation of 1.1\arcsec , therefore not visible  in Fig~\ref{IRDIS_IFS_RDI}.

\begin{figure*}
\centering
\includegraphics[trim={1.7cm 13cm 2cm 1.7cm},clip,width=\textwidth]{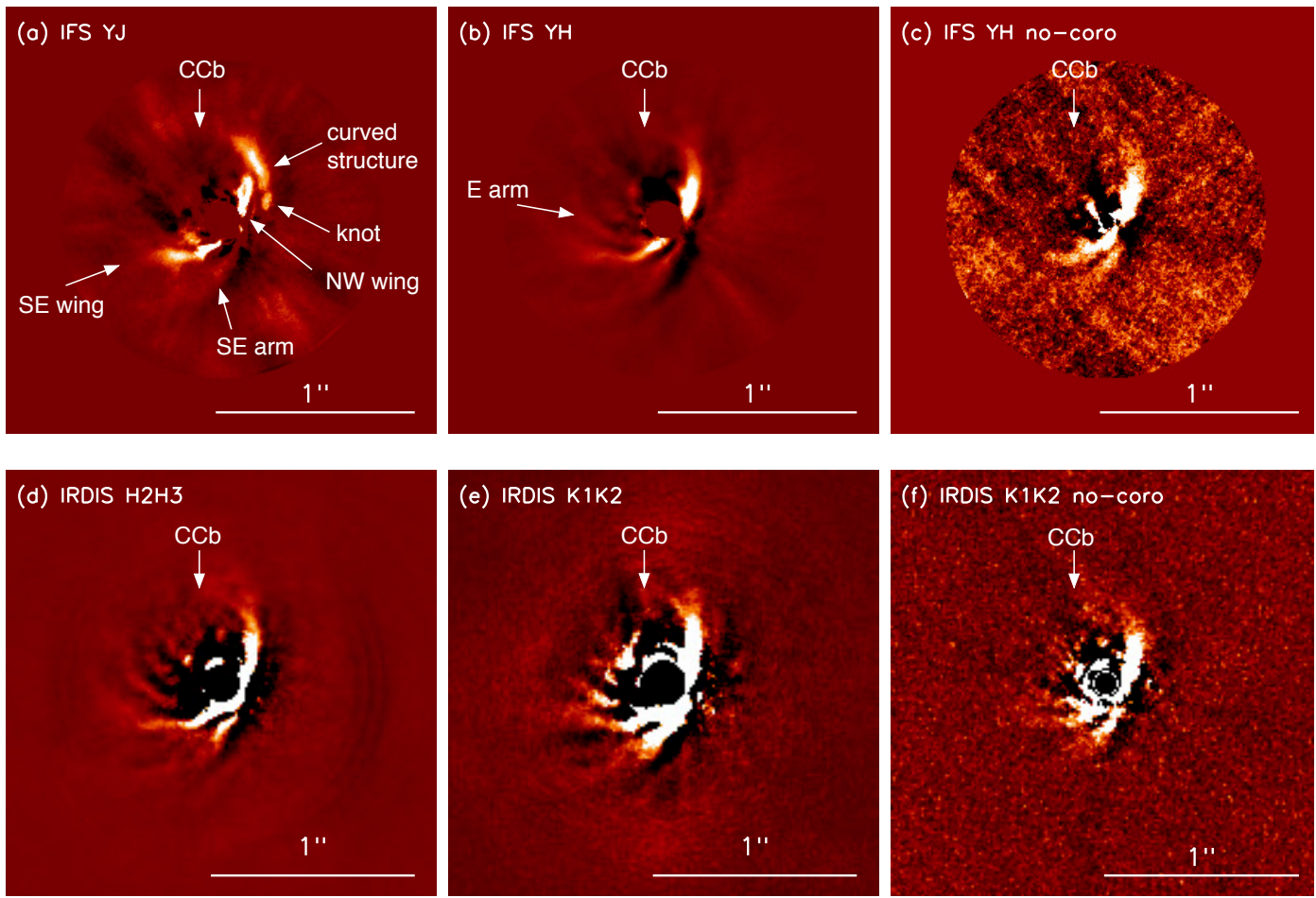}
\caption{ADI images obtained for HD 100546. Top: IFS cADI images: (a) YJ from June 15; (b) YH from May15; (c) JH no coronagraphic image from Apr 16; Bottom: IRDIS TLOCI images (d) H2H3 from June 2015; (e) K1K2 from May 2015; (f) K1K2 no coronographic image from Feb 17. In all the images North is up, Est is left.}
\label{IFSIRDIS_simpleADI}
\end{figure*} 

Figure~\ref{IFSIRDIS_simpleADI} shows the results of the cADI analysis of the SHINE observations of HD~100546, where several disk structures are well visible. 

In the first row, we report for the first time SPHERE/IFS images of the disk-bearing HD~100546. YJ collapsed image (Fig.~\ref{IFSIRDIS_simpleADI}a) shows the \textit{two wings} described in \cite{2016Garufi} with a higher angular resolution that allows to detect an additional curved structure in the NW side, between 284\degree\ and 350\degree, at separation of about 0.3\arcsec\ and a southern spot. IFS also recovered the SE fainter arm. The IFS YH collapsed image (see Fig.~\ref{IFSIRDIS_simpleADI}b) reveals a single wider northern wing that barely reaches the expected position of CCb. The prominent southern wing extends up to 0.7\arcsec\ and PA=90\degree. Two additional spiral arm-like features are also clearly visible. Referring to \cite{follette2017}, that provide a clear description of the innermost structure of HD~100546, these two arms may be identified with \textit{"S4"} and \textit{"S6"}; the southern one corresponds to the SE arm. In our image, an additional arm that lies above the southern wing, and starts from it, seems to extend in the East direction up to $\sim$0.6\arcsec\  and PA=90\degree (E arm in Fig.~\ref{IFSIRDIS_simpleADI}b) and can be properly reproduced with a logarithmic spiral with a large pitch angle ($\sim50$\degree). There is also a hint of the {\it "S2"} spiral structure observed by \cite{follette2017}, E to the northern wing. No bright PSF-like knots are recovered along the arms at these wavelengths. We confirm the presence of a dark area just offset from the wings in the West direction, up to $\sim0.3$\arcsec, also detected in the RDI, where the disk shape is less altered thanks to a negligible self subtraction. This dark area is also present in the IFS YJ images where it looks less deep. In Fig.~\ref{IFSIRDIS_simpleADI}c the non-coronagraphic YH collapsed image clearly reveals the two bright wings and the SE arm and it appears clear that the distribution of light is uninterrupted between the wings, with only a small depletion along the minor axis that is likely an artifact of the ADI analysis: the two wings appear then as a unique structure, symmetric to the minor axis, whose central part lies very close to the star. The darker area in the West is still visible.

In the bottom row of Fig.~\ref{IFSIRDIS_simpleADI} we present IRDIS TLOCI images. In Fig.~\ref{IFSIRDIS_simpleADI}d and \ref{IFSIRDIS_simpleADI}e, we show the median H2H3 and  K1K2 images, respectively. The two wings, the SE arm, the East arm and the dark area appear clearly visible in both images, but show some differences with respect to shorter wavelengths. A diffuse signal is visible on the top of the northern wing in H2H3 and also in K1K2, but is less luminous. This is located close to the expected position for CCb \citep{2015Quanz,2015Currie}. The East arm extends up to $r$=0.7\arcsec\ and PA=64\degree\ in the H2H3 filter pair but only to $r\sim$0.4\arcsec\ and PA$\sim90$\degree\ in K1K2. In the H2H3 image, a second dark area is visible into the west side starting at $\sim0.6$\arcsec\ till $\sim0.9$\arcsec. At larger separations, at least two spiral arms are visible in the south, ranging from  $\sim1.1$\arcsec\ to $\sim2.3$\arcsec. 

We notice that the SPHERE data were obtained under worst seeing conditions than GPI ones. The biggest difference in the disk structures concerns the dimension of the "S4" spiral \citep{follette2017} which looks more extended in the GPI observations than in the SPHERE data. Moreover, a few differences between GPI and SPHERE observations can be noted on the extended structures around the location of the candidate companion CCb. We will discuss this in Sect. 4.

\subsubsection*{Polarimetric Images}
\label{sec:PDIresults}
\begin{figure*}
\centering
\includegraphics[width=\textwidth]{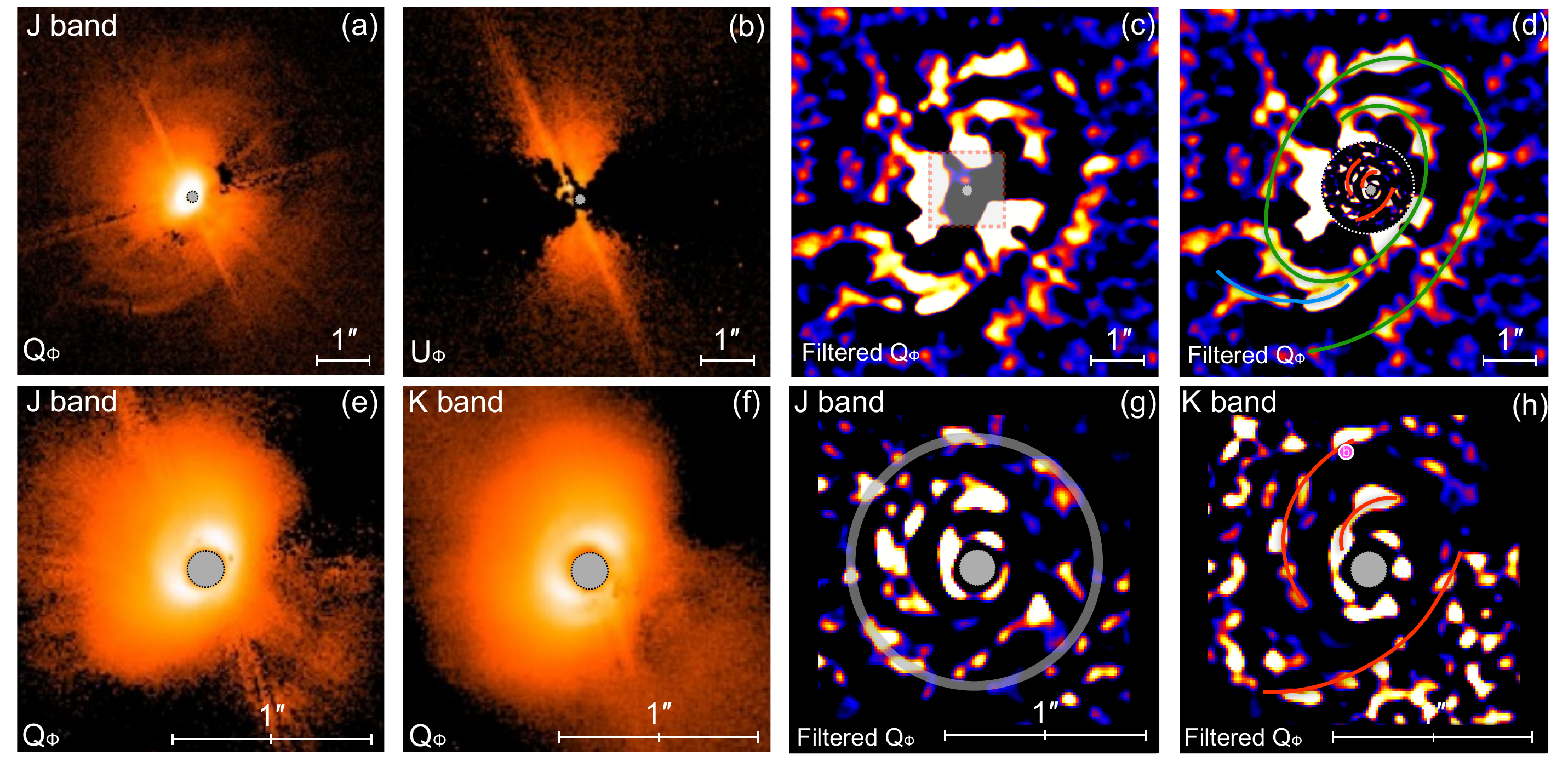}
\caption{Polarized light images of HD100546 from SPHERE/IRDIS. (a): The $Q_{\phi}$ image in the J band. (b): The $U_{\phi}$ image in the J band. (c): Unsharp masking of the $Q_{\phi}$ image in the J band (see text). The inner box defines the region displayed in the bottom row. (d): Labeled version of (c). The inner inset circle is from (g). (e): Inner detail of the $Q_{\phi}$ image in the J band. (f): Inner detail of the $Q_{\phi}$ image in the K band. (g): Unsharp masking (see text) of (e). The grey circle indicates the IRDIS control radius at $r\approx 20 \times \lambda /D$ (see text). (h): Unsharp masking (see text) of (f). Features visible from both (g) and (h) are labeled. The purple dot indicates the location of CCb from \citet{2013Quanz1}. In all images, the central star is in the middle of the grey circle, symbolizing the instrument coronagraphic mask. The logarithmic color stretch is  relatively arbitrary and refers to positive values, except in (a) and (b) where it is the same. North is up, East is left.}
\label{PDI}
\end{figure*}

The polarimetric images resulting from the data reduction described in Sect. \ref{sec:PDIobs} are shown in Fig.~\ref{PDI}. We found that the outer disk ($>1$\arcsec) is well imaged in the short exposure in the J band. The K-band image is instead affected by thermal background in the detector, which prevents us to obtain similarly good image at larger radii as in the J-band.

The $Q_{\phi}$ image of the whole system is shown in Fig.~\ref{PDI}a, whereas the respective $U_{\phi}$ is shown in Fig.~\ref{PDI}b. Strong signal is detected from both images. In particular, the $U_{\phi}$ image shows positive values to North and South and negative values to East and West. Four prominent stellar spikes are left from the data reduction in both images. A possible explanation for these artifacts is related to the bright ($J$=$6.42$ mag) central star polarization degree that makes the spikes not perfectly cancelled throughout the process of subtraction of the beams with different polarization states. There is indeed a compact polarized inner ring (0.24-0.7 au, \citealt{2014Panic}, unresolved from the star in our images) that can contribute, even if the near-IR excess originating from that region is not particularly large for this source and this effect is not appreciable in other stars with similar sub-au disks. 

The circumstellar disk is clearly visible in Fig.~\ref{PDI}a. To image a larger flux range, the images are shown in logarithmic scale. In the image, a number of wrapped arms at radii $1.5$\arcsec -3\arcsec\ from the central star are visible at all azimuthal angles. In particular, we recover the arms to South imaged by \citet{boccaletti2013} and \citet{2016Garufi} and those to North visible from the Hubble Space Telescope (HST) image by \citet{2007Ardila}. 

To highlight elusive disk features from Fig.~\ref{PDI}a, we applied an unsharp masking to the image similarly to what is described by \citet{2016Garufi}. This technique consists in subtracting a smoothed version of the original image to the image itself. The result of the unsharp masking, obtained by smoothing Fig.~\ref{PDI}a by $\sim 10$ times the angular resolution of the observations, is shown in Fig.~\ref{PDI}c. Looking at this image, it is possible that the majority of the structures visible from Fig.~\ref{PDI}a are part of a unique arm which is wrapping for at least $540$\degree, as indicated by the green line in Fig.~\ref{PDI}d and as already noticed by \cite{garufi2017}. Another bright arm visible to South seems to have a similar origin but to extend Eastward with a larger aperture angle (see the cyan line of Fig.~\ref{PDI}d).   

In the inner zone of the disk, inside $\sim 1$\arcsec, the polarized flux is dramatically higher, even though two regions with reduced flux (appearing black in the image) are visible at P.A.$\sim 160$\degree\ and P.A.$\sim 300$\degree\ (Fig.~\ref{PDI}a), similarly to what found by \citet{2011Quanz} and \citet{avenhaus2014}.

Fig.~\ref{PDI}e and of Fig.~\ref{PDI}f show the zoomed $Q_\phi$ image in J and K band of the innermost disk region. Two bright lobes are visible to the SE and to the NW, with the former being brighter than the latter. Inward of these lobes, the disk cavity is marginally visible just outside of the instrument coronagraphic mask. Similarly to Fig.~\ref{PDI}a, disk features are not easily recognizable. Therefore, we also apply an unsharp masking to these images, with smoothing by $\sim 6 \times$FWHM, which results in Fig.~\ref{PDI}g and Fig.~\ref{PDI}h respectively. In both images, a number of possible features are visible. Among them, we only give credit to those that are persistent across unsharp masking procedures (i.e., by varying the smoothing factor) and, more importantly, across wavebands. Features that are radially distributed at $r \approx 20 \times \lambda / D$ ($\sim60$ au) are masked out in Fig.~\ref{PDI}g, because this region corresponds to the SPHERE control radius (see Fig.~\ref{PDI}g). Features present in both the J and K band images are labeled in red in Fig.~\ref{PDI}h. The morphology of the identified structures resembles the shape of the arms at larger radii and is consistent with the known disk geometry. The innermost of these features was also detected by \citet{avenhaus2014} and \citet{2016Garufi}, whereas the others were not. Interestingly, CCb as from \citet{2013Quanz1}  lies in correspondence of one of these possible features. From this data set, we cannot infer whether a spatial connection between the inner arms (in red) and the outer arm (in green) exists. 

\subsection{Comparison between ADI and PDI}
\label{sec:conf}
\begin{figure*}
\centering
\includegraphics[trim=0cm 2cm 0cm 0cm, clip, width=0.8\textwidth]{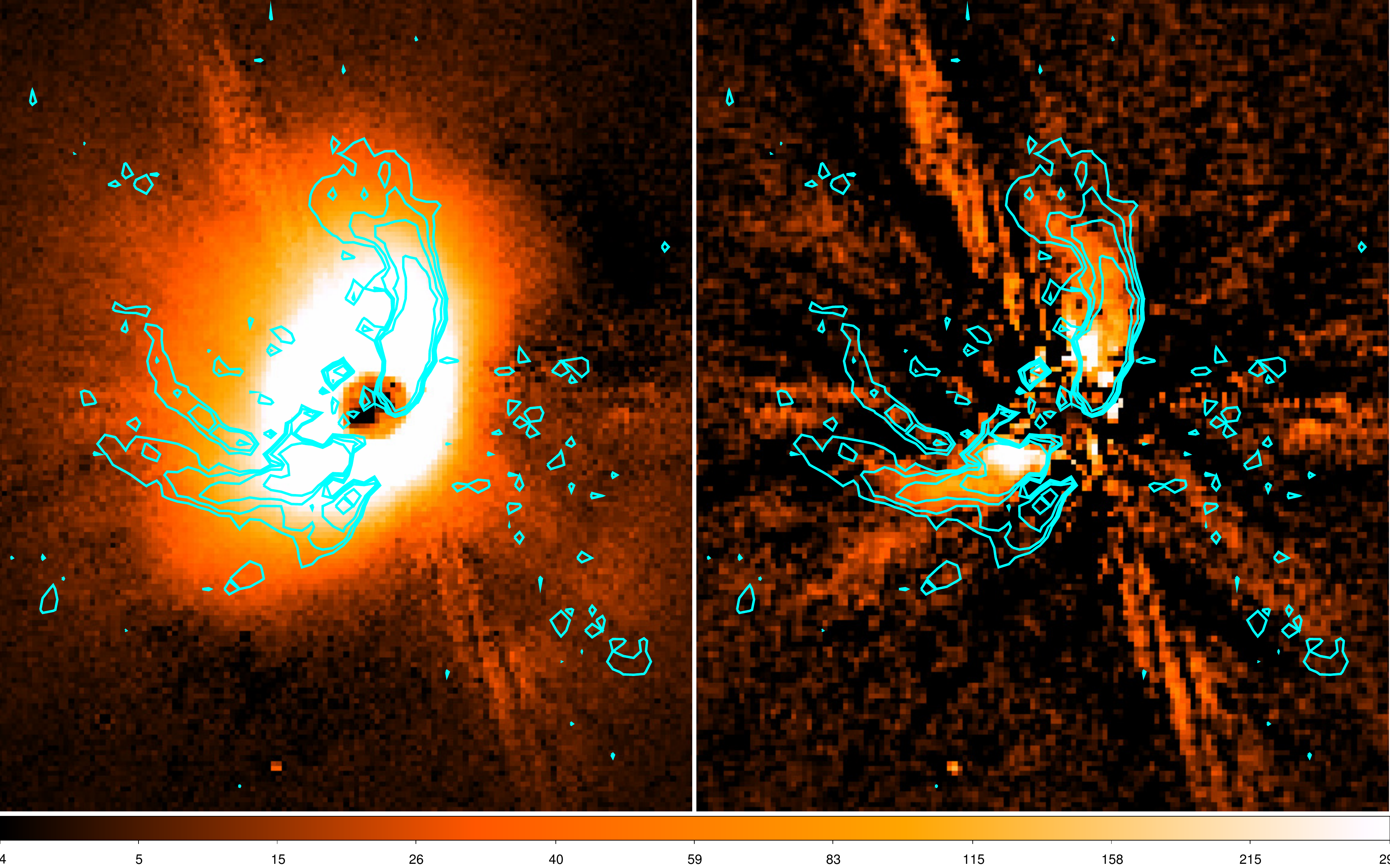} 
\caption{IRDIS PDI J band $Q_\phi$ image (left) and the results of a simulated cADI analysis on this image. Overplotted in cyan the isophotes of IRDIS cADI H2 observations.}
\label{IRDISPDIADI}
\end{figure*}

While the southern spiral arms of the outer disk ($>1$\arcsec) in the IRDIS H and K images are well detected also in the IRDIS PDI J (Fig.~\ref{IFSIRDIS_simpleADI}d-e and Fig.~\ref{PDI}a), the central part of the PDI data do not show any axis-asymmetric structure. Using the unsharp masking technique on PDI data, in the innermost region we found out three spiral arms (Fig.~\ref{PDI}h) that have no counterpart in the intensity image. To compare ADI results with the PDI ones, we have then performed pseudo-ADI on the IRDIS DPI J-band images as follows: we use the parallactic angle values of the three different IRDIFS\_EXT  observations to simulate a simple cADI analysis based on the IRDIS DPI image. The results are given in Fig.~\ref{IRDISPDIADI}: the simulation generates an image that is in better agreement with the IRDIS cADI images. On the other hand, PDI images are directly comparable with RDI images. These two methods allow to better study the light distribution in the disk without self subtraction effects.

\subsection{Disk spectrum}
\label{sec:spectrum}

\begin{figure}
\centering
\includegraphics[width=\columnwidth]{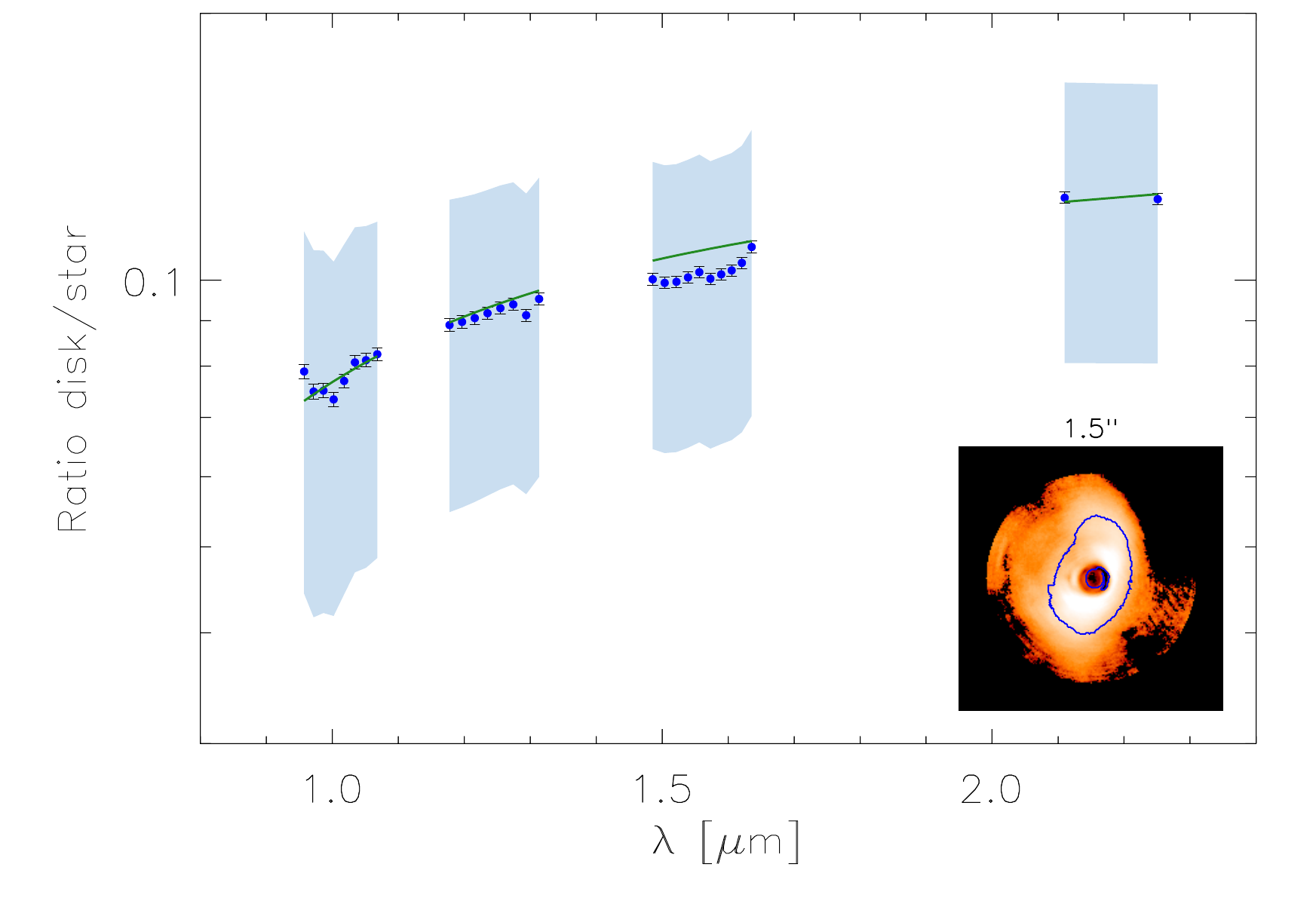}
\caption{Disk mean spectrum of HD 100546 along the wings. The blue dots and the shaded regions refers to the median value and its 3$\sigma$ uncertainty. The green line represent the best fit obtained with light reflected by dust with constant albedo $A=0.65$. In the bottom-right inset, the area selected for the spectrum extraction on the RDI images is delimited in blue.}
\label{disk_spectrum}
\end{figure}

In general, disk flux in the NIR is dominated by scattered light, while the brightness of a young planet is dominated by thermal emission. Therefore, to distinguish between a self shining planet and the disk we studied the total spectrum of the HD~100546 disk. Previous attempts to derive the spectrum of the disk of HD~100546 were done by \cite{2013Mulders} using HST data, and by \cite{stolker2016} using data from NACO at VLT.

In order to derive the disk spectrum, we estimated the ratio between the total disk flux and the stellar one. We used the RDI datacube obtained with IFS in May 2015, where self subtraction is less aggressive. The target to reference intensity ratio is measured in a specific region of the image, defined as the pixels of the wavelength-combined RDI image with a value above a determined threshold. This procedure yielded an elliptical region with deprojected radius $\sim40$~au, without the central region that corresponds to the coronagraph, as shown in the small inset in Fig.~\ref{disk_spectrum}. An adequately rescaled mask, in order to sample the same area of the disk, was also applied to the IRDIS RDI frames. Several reference stars were tested to confirm that the choice of the HD~100546 data set and the reference star do not affect the final contrast spectrum (all data sets considered have correlation coefficient $> 0.98$ with that of HD~100546). 

The relative flux spectrum is shown in Fig.~\ref{disk_spectrum}. Since we are interested in the disk albedo, the flux calibration of the disk is based on broad-band photometry. The uncertainties on each wavelength (blue bars) were estimated as the standard deviation of the mean of the spectra for individual pixels. Moreover we also show the intensity $1\sigma$ range over the population of individual pixels (light blue area). The regions between 1.08-1.15 $\rm{\mu m}$ and 1.35-1.48 $\rm{\mu m}$ are affected by strong water absorption due to the Earth atmosphere that cannot be perfectly recovered during the RDI process and were therefore removed.

The difference in contrast magnitude is close to 0 between H and K bands (K-H $\sim 0.03$, in agreement with similar analysis presented by \cite{avenhaus2014} exploiting NACO $H$, $K_s$ and $L'$ filters), and the spectrum appears featureless, with the disk reflecting about 8\% the stellar light in the Y band and about 11\% in K band. This is very similar to the spectrum obtained by \cite{2013Mulders}. We tried to represent this spectral distribution assuming light reflected by dust with a constant albedo. \cite{2013Mulders} suggested that in the near-infrared regime, light scattered off the HD~100546 outer disk surface is not only scattered stellar light, but the contribution of the light reprocessed by the inner disk \cite[0.24-0.7 au,][]{2014Panic} is not negligible. Therefore, we took into account a possible reddening due to the presence of the inner disk that can absorb stellar light. This was done following \cite{cardelli1989}. The free parameters are then reddening and the (total) disk reflectivity. The first is related via extinction ($\mathrm{A_V}$) and reddening ($\mathrm{R_V}$) to the slope of the spectrum, the latter is the combination of the albedo and the disk emitting area defined above\footnote{Using the model derived in Sec. \ref{sec:disk_model}, the disk height at 40~au corresponds to 7.7\degree\ above the disk mid-plane.}. The best solution gives $\mathrm{A_V}$=$2.11$ and albedo $A$=$0.96$ that is very high and not credible. An alternative more palatable explanation is that the albedo of dust is not constant with wavelength. A similar explanation was already suggested by \cite{2013Mulders} and \cite{stolker2016}. A variation of the scattering efficiency from 0.43 at 1 $\mu$m up to 0.63 in the K-band may explain the observations. This points toward rather large particle sizes.

The reddened spectrum we derive was also seen in HD~100546 HST data and was explained by the effect of forward scattering of micron-sized particles \citep{2013Mulders}. Further analysis based on the PDI data confirmed the presence of these particles, determining the phase function \citep{stolker2016}.

\subsection{Disk geometrical model}
\label{sec:disk_model}

The disk structure of HD~100546 has been investigated in different wavelengths regimes. ALMA observations suggest that the millimeter-sized grains are located in two rings: a compact ring centered at 26 au with a width of 21\,au, and an outer ring with a width of $75 \pm 3$\,au centered at $190\pm3$\,au \citep{walsh2014}. The same two-rings structure was suggested by \citet{2014Panic}, who used MIDI/VLTI data to probe micron-sized grains. They also came to the conclusion that there is an additional inner disc that extends no farther than 0.7\,au from the star. The gap is about 10\,au wide and free of detectable mid-infrared emission. Much larger dust, rocks, and planetesimals are not efficient H-band emitters, and our data do not exclude their presence inside this gap. Analysis of the SED of HD~100546 shows that the radial extent of the gas is $\sim400$\,au \citep[e.g.,][]{benisty2010}. The presence of an extended disk is clearly indicated also by scattered light, NIR and sub-millimeter imaging \citep{2000Pantin, 2007Ardila, 2013Quanz1, avenhaus2014,2014Currie, 2016Garufi, follette2017}.\\

To constrain some of the HD~100546 disk properties we built a simple geometrical model based on analytic functions, that can describe the inner part of the system, between 10 and 200 au. This model is similar to that described by \cite{stolker2016}. The basic assumption of this model is that the disk photometry can be described by light scattering on a single surface. In our model, the scattering function for total intensity images is  described by a two components  Henyey-Greenstein (HG) function \citep{1941ApJ....93...70H}  with coefficients as defined in \cite{milli2017} for HR~4796. For PDI data we used the function given by \cite{stolker2016}, $cfr.$ Eq. (8). The light from the star is reflected by the optically thick disk surface, that lies above the disk mid plane as described by the power law: 
\begin{equation}
H = c \left(\frac{r}{r_0}\right)^{b} \mathrm{\quad[au]},
\end{equation}
where $r_0$ corresponds to 1~au and $c$ is in au. Gaps present in the disk are modelled by decreasing the disk height $H$ to zero au. This does not mean that the gaps are really empty, but simply that we cannot model emission from these regions with our approximation. Assuming the disk geometry as found from the MIDI data, which trace micron-sized grains, and the ALMA data, which trace millimeter-sized grains, we have also assumed for the scattered light traced by SPHERE the same disk geometry, being aware that different wavelengths probe of course different grain size and that different grain size may occupy different physical location in the disk. Therefore we set the rings range between $\sim$10 and $\sim$40\,au (here after ring 1) and then from $\sim$150 to $\sim$230\,au (hereafter ring 2). 

The thin inner ring  (hereafter ring 0) extending between 0.24 and 0.7\,au is behind the SPHERE coronagraph and therefore it is not included. The positions of the inner and the outer edge of the ring 1 are well constrained by the observed radial profiles: the inner edge must be at least partly  behind the coronagraph, otherwise the disk wall will be clearly visible on the semi minor axis radial profile on our images, while the outer edge is visible on the semi major axis. The flaring parameter $b$ is derived by the flux ratio between the peak of the near and the far side of the semi minor axis while the scale factor $c$ impacts the shadowing effect of the disk. 

\begin{figure*}
\centering
\begin{tabular}{cc}
\includegraphics[trim=0.2cm 0.3cm 0.8cm 0.3cm, clip,width=0.9\columnwidth]{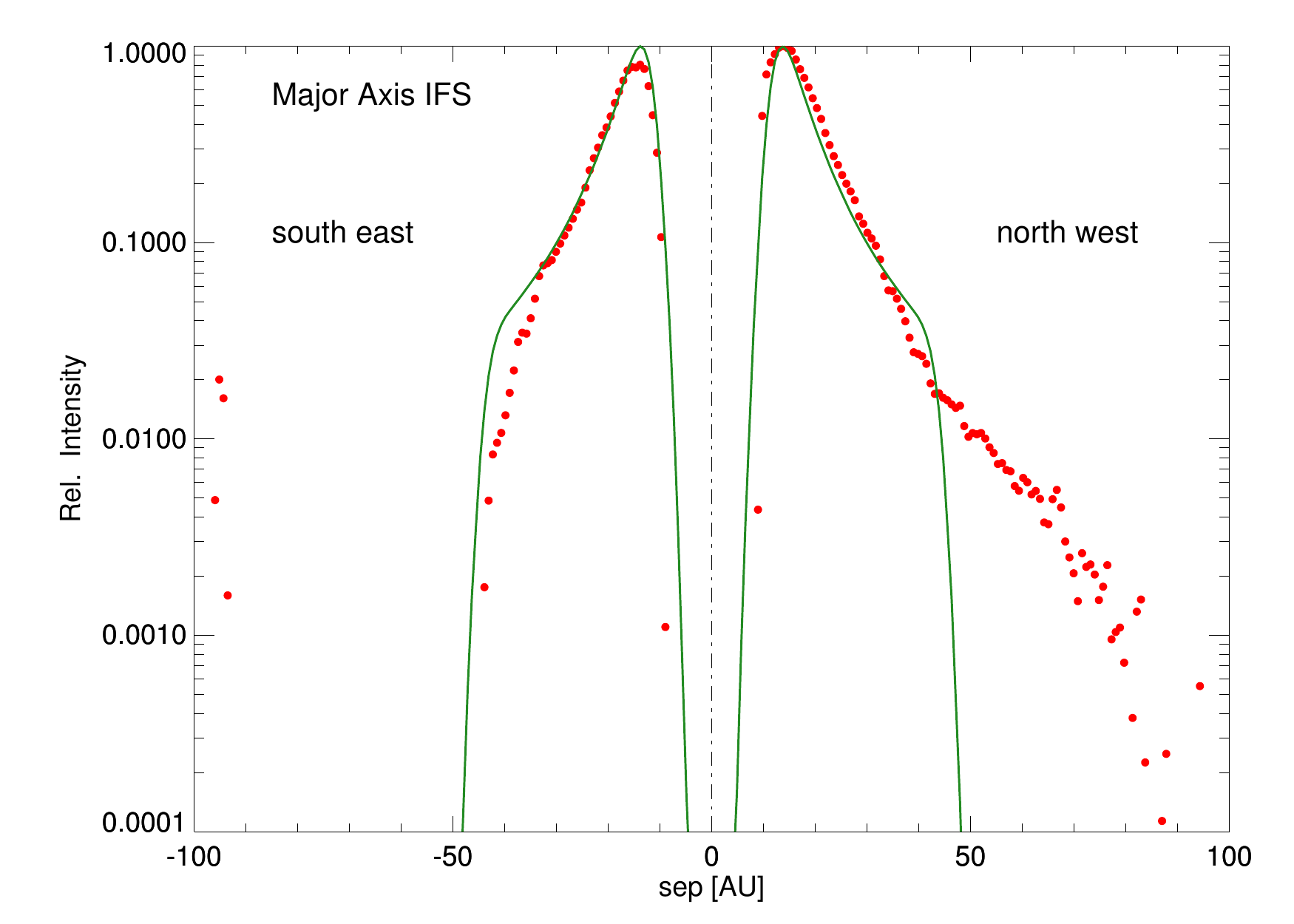}&
\includegraphics[trim=0.2cm 0.3cm 0.8cm 0.3cm, clip,width=0.9\columnwidth]{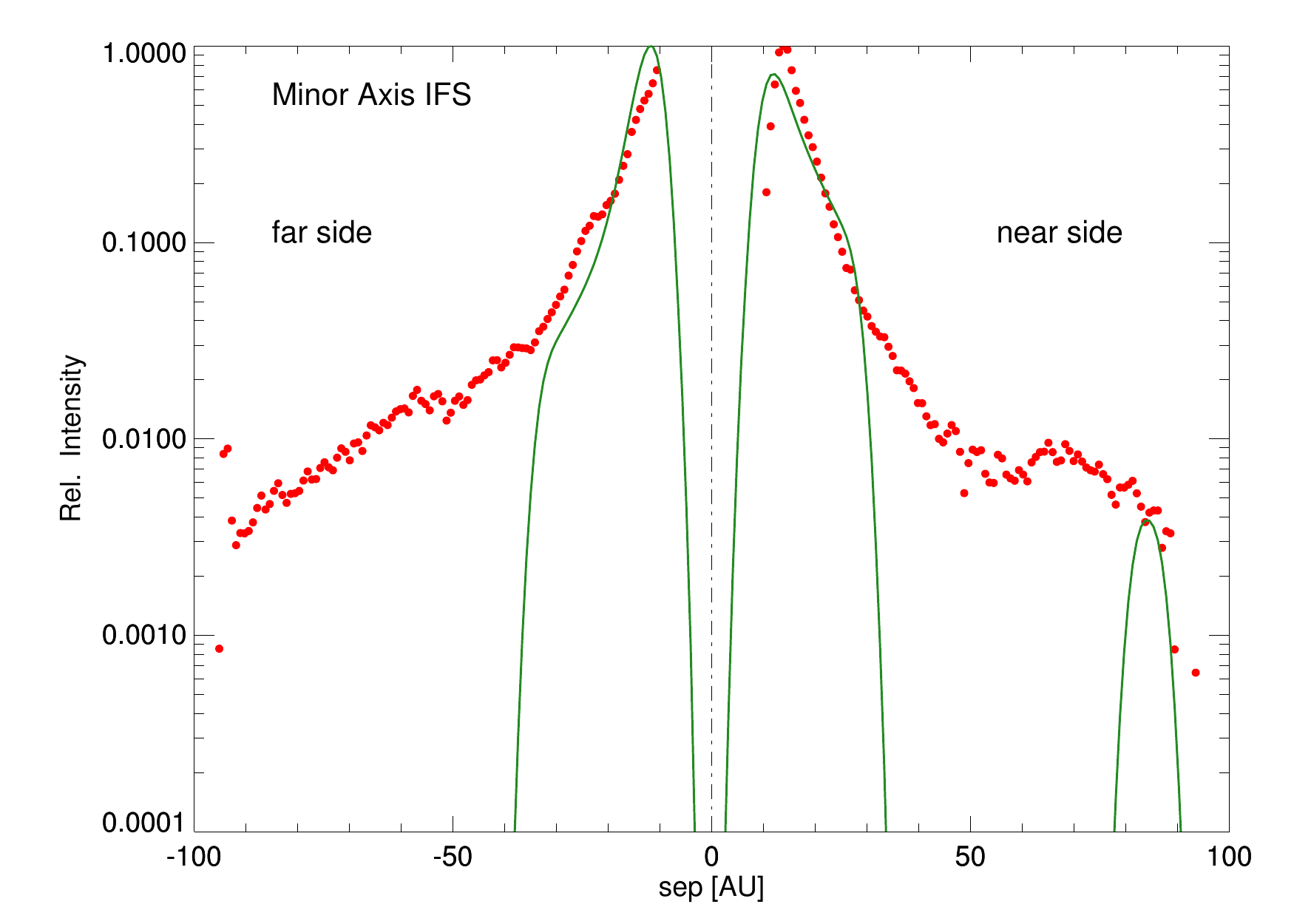}\\
\includegraphics[trim=0.2cm 0.3cm 0.8cm 0.3cm, clip,width=0.9\columnwidth]{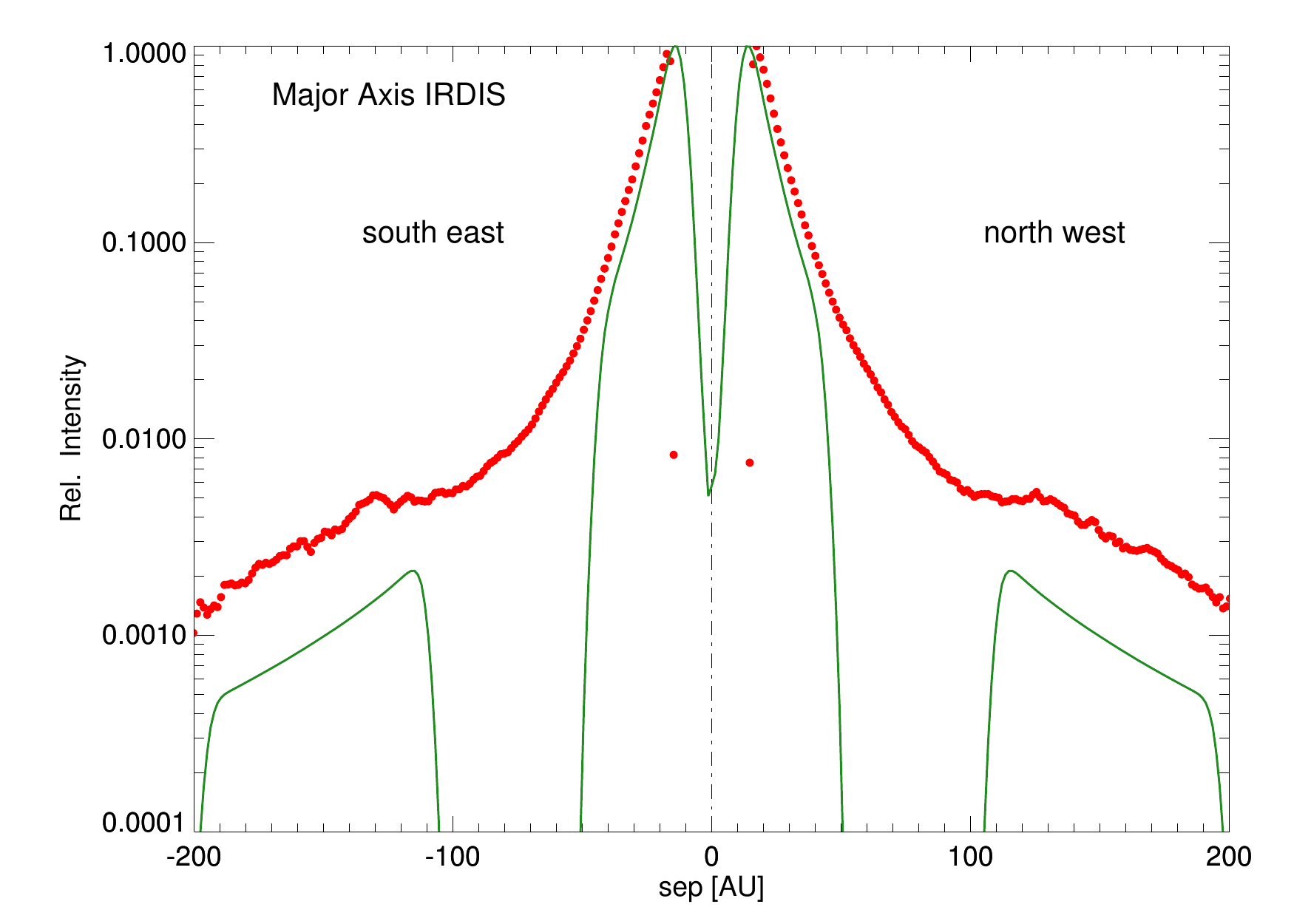}&
\includegraphics[trim=0.2cm 0.3cm 0.8cm 0.3cm, clip,width=0.9\columnwidth]{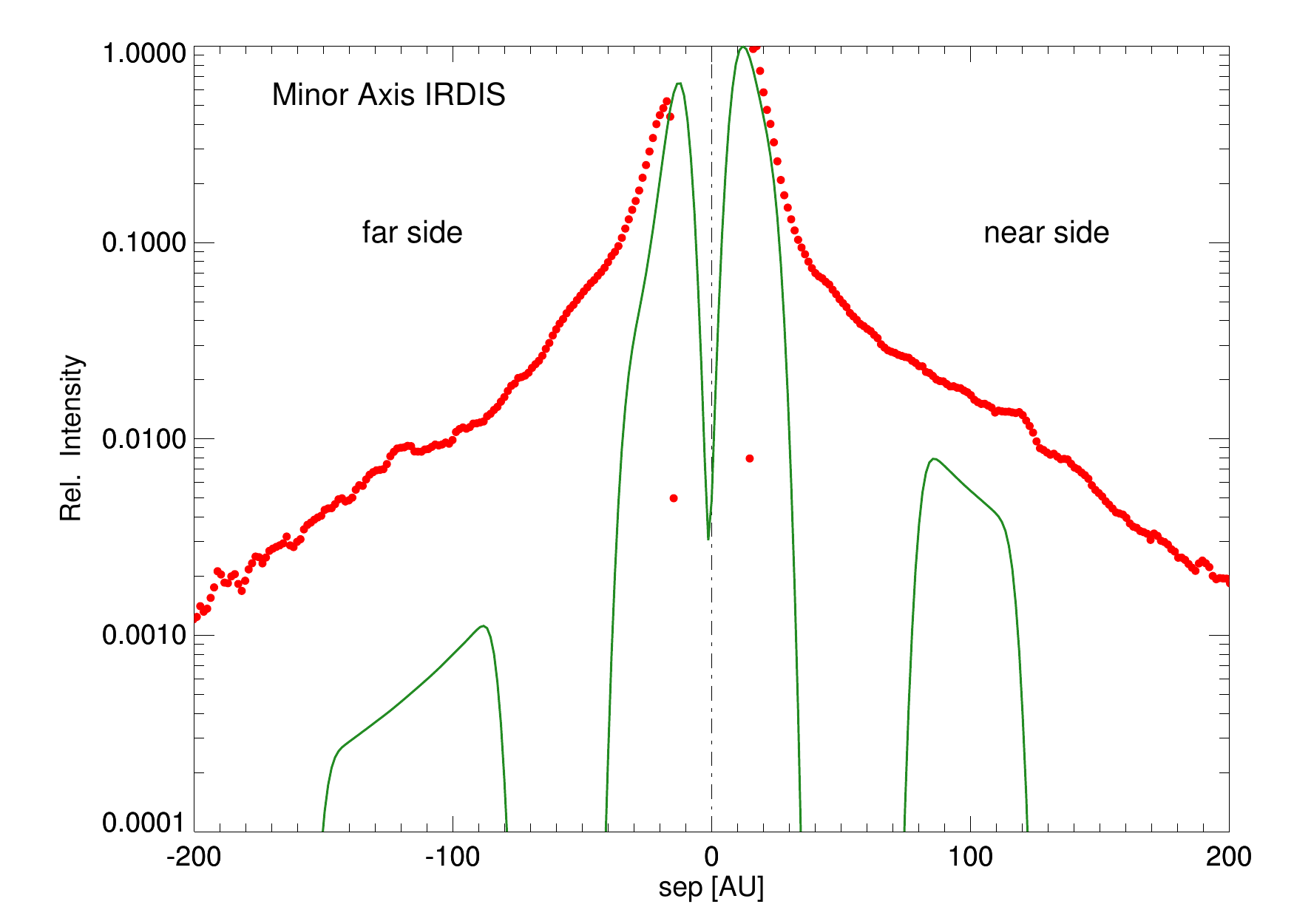}\\
\end{tabular}
\caption{Radial profile along  the semi-major and the semi-minor axis of HD~100546 RDI images (red dots) compared with the best fit model (green line) obtained for IFS and IRDIS.}
\label{modelRDI}
\end{figure*}

\begin{figure*}
\centering
\begin{tabular}{cc} 
\includegraphics[trim=0.2cm 0.3cm 0.8cm 0.3cm, clip,width=0.9\columnwidth]{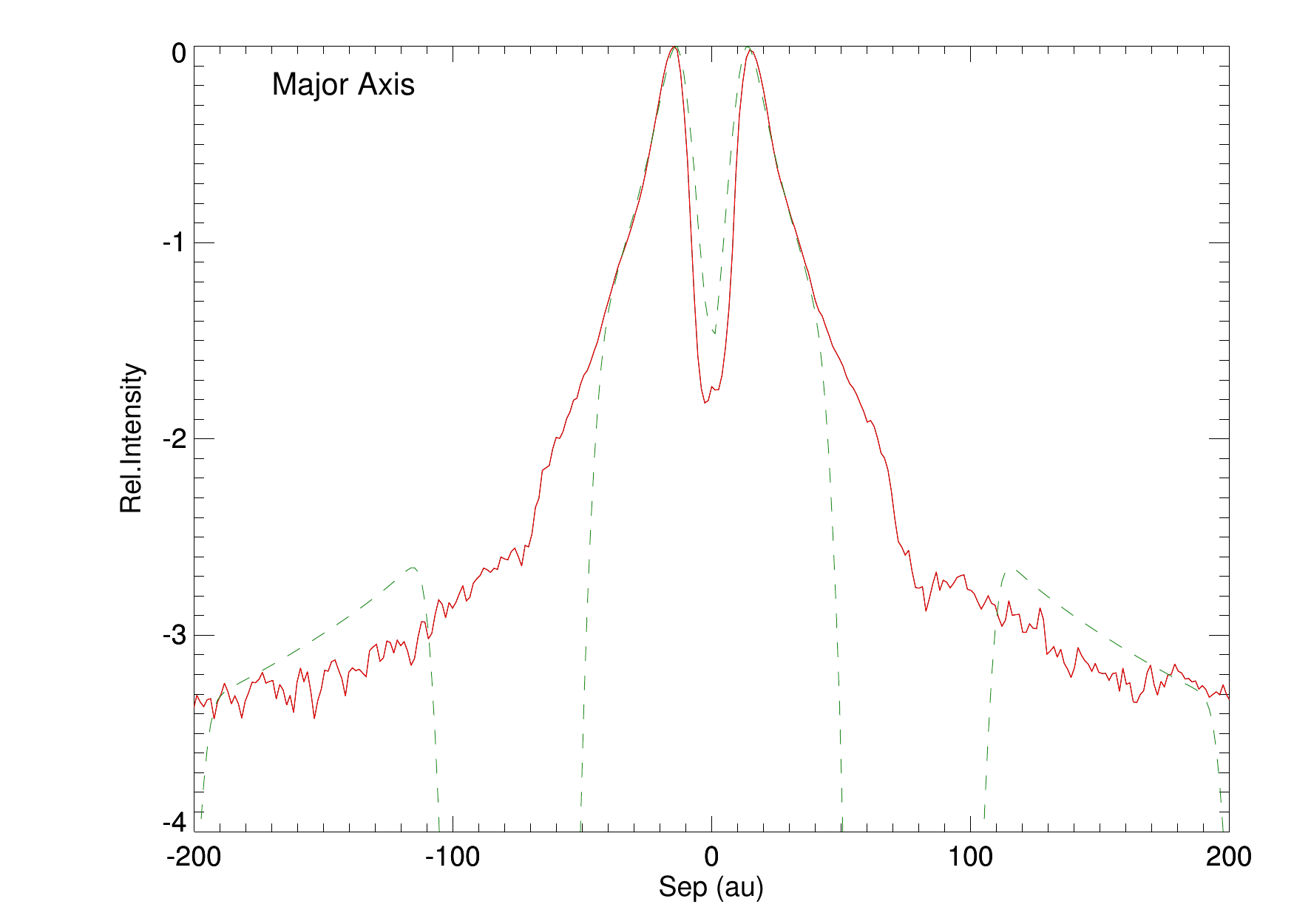}&
\includegraphics[trim=0.2cm 0.3cm 0.8cm 0.5cm, clip,width=0.9\columnwidth]{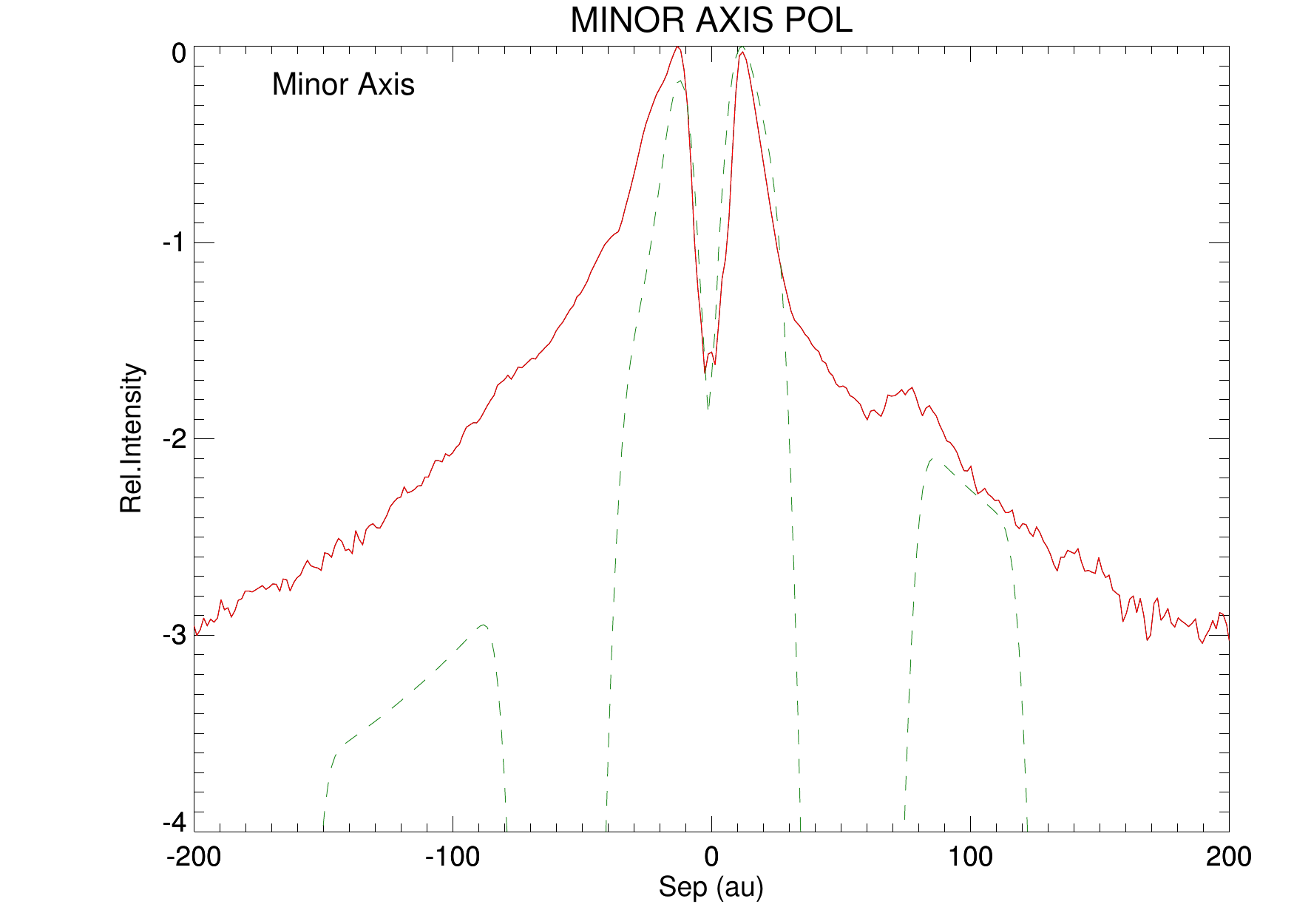}\\
\end{tabular}
\caption{Radial profile along the semi-major and the semi-minor axis oh HD 100546 PDI images (continuum line) compared with the best fit model (dashed line) IRDIS.}
\label{modelPDI}
\end{figure*}

\begin{figure*}
\centering
\includegraphics[trim=0cm 2cm 0cm 0cm, clip, width=\textwidth]{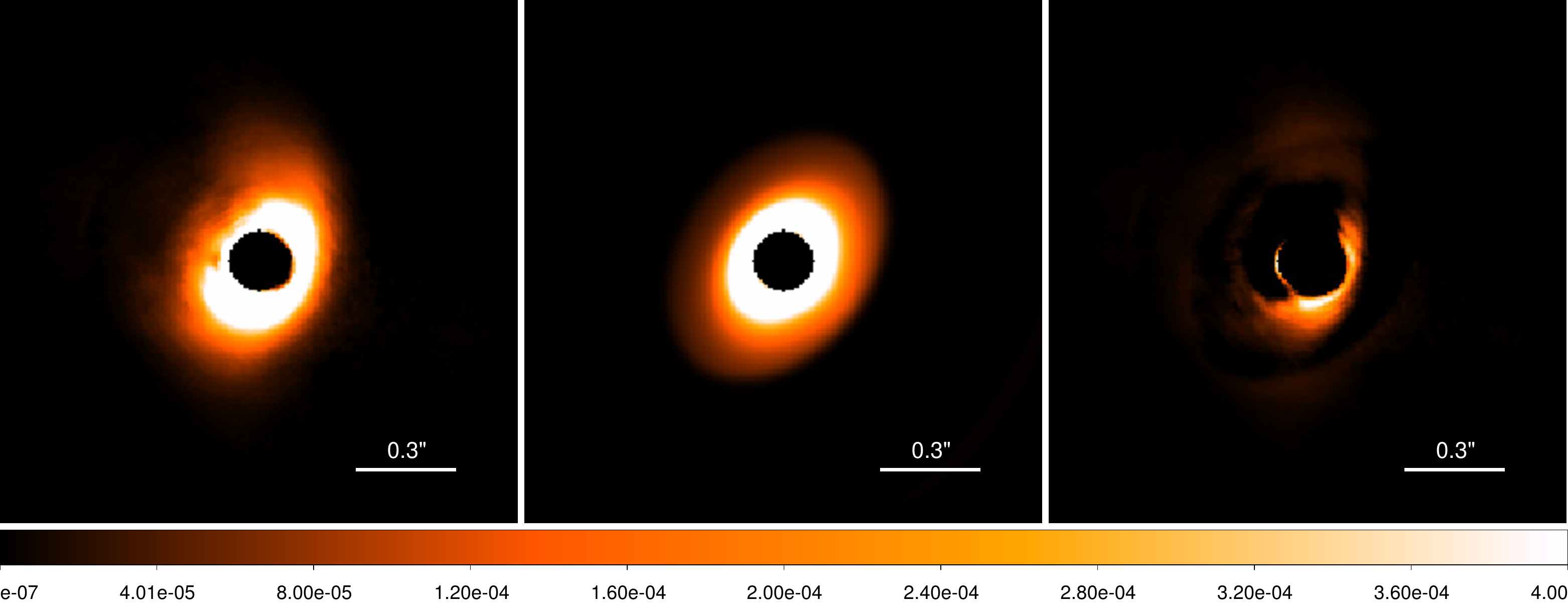} 
\caption{From left to right, IFS  RDI image, model and residuals. Image are displayed with the same linear color bar from $6\times 10^{-3}$ dex to 1 dex. North is up, East is left.}
\label{modelRDIimg}
\end{figure*}

We first compared our model to RDI images; we did it for IFS and applied the same model also to IRDIS. We found out that a disk with $b$=$1.08$ and $c$=$0.13$ best reproduces the HD~100546 radial profiles (Fig. \ref{modelRDI}), quite similar to the $\tau$=1 disk surface for $H$ and $K$ band derived by \cite{stolker2016}. The disk rings 1 and 2 extend between $\sim12$~au and $\sim44$~au and between $\sim110$~au and $250$~au respectively, consistent with previous results based on ALMA observations \citep[e.g.][]{walsh2014}. The model for the outer disk is however fainter than observed, suggesting a larger flaring at large separations, though it should be considered that the RDI images are not very accurate at very high contrast levels. However, this model is also in quite good agreement with the IRDIS PDI images in broad J band as shown in Fig. \ref{modelPDI}. We notice that the inner edge of the ring 2 and the outer edge of the ring 1 might correspond to resonances 3:2 and 1:2 with a  hypothetical massive object located  at 70\,au radius orbit, that is not far from the observed location of CCb (see next Section).

The resulting RDI images are shown in Fig.~\ref{modelRDIimg}. The residuals clearly show the presence of the two wings, that are even more evident when applying to this model the ADI method. To show this, we constructed a data cube (x, y, lambda, time) made of the sum of the disk model and of the reference image used in our RDI analysis, that is the data set of HD~95086. This datacube was then processed by the same ADI routine used for the original HD 100546 datacube. In this case, the two bright wings visible in all the images in Fig.~\ref{IFSIRDIS_simpleADI} can be reproduced only by assuming a much thicker disk, with a value of $H\sim30$\,au at $r\sim40$\,au. This is because the bright wings visible in the ADI images, that correspond to the edge of the disk, are described by an ellipse whose center is far from the star. 

A similar result has been found for several disks, a classical example being HD~97048 \citep{ginski2016}. As explained in their Fig.~5, disk images that may be described by off-center ellipses indicate the presence of material located well above the disk mid-plane. This was also demonstrated by recent hydrodynamical simulations performed by \cite{dong2016} coupled with simple radiative transfer models. These simulations demonstrate that a giant planet can open a gap in a disk creating a ring that, when seen at a intermediate viewing angle, and reinforced by the ADI process as explained above, appears as two pseudo-arms placed symmetrically around the minor axis winding in opposite directions. They conclude that these simulations can describe HD~100546 as well. They also predicted the presence of a "dark lane" parallel to these disk pseudo-arms that is due to the self shadowing by the disk. In the case of HD~100546, the elliptical light distribution may indicate the presence of a ring of material at $\sim$40 au, close to the outer edge of the intermediate disk, but at an height of $\sim30$\,au on the disk plane. This is about four times the value obtained with Eq. 2, using the best value we obtained for $b$ and $c$. The origin of this ring is not well clear; we remind however that the edge of the intermediate disk can be explained by a 2:1 resonance with a hypothetical massive objects located at about 70\,au. 

In order to agree with PDI and RDI data, we have to assume that this material is optically thin, contributing only marginally to the total emission from the disk. However, the higher (about a factor 25) sensitivity provided by ADI allows its detection. 

We note that a multi-rings configuration similar to that observed in HD~100546 is also seen around other objects. We mention in particular HD~141569A, where rings and outer spiral arms were observed with HST in the visible \citep{augereau1999, mouillet2001,clampin2003}, and in the near-IR with NICI \citep{biller2015,mazoyer2016} and SPHERE \citep{perrot2016}. Moreover, the face-on disk surrounding TW Hya \citep{vanboekel2017} presents three rings and three gaps within $\sim$2\arcsec\ from the central star. These features were identified using optical and near-infrared scattered light surface brightness distribution, observed with SPHERE. Multiple rings were also observed in the inclined system RX J1615.3-3255 \citep{deboer2016} combining both visible polarimetric images from ZIMPOL with IRDIS and IFS in scattered light. 

Finally, we notice that  planet sculpting a cavity is not a unique explanation to the gaps. In addition some models with planets located outside of the gap can reproduce the gaps, or a single planet can open, under particular circumstances, multiple gaps \citep{dong2017}.

\section{The candidate planets}
\label{sec:planets}

\subsection{Detection limits}

\begin{figure}
\centering
\includegraphics[width=\columnwidth]{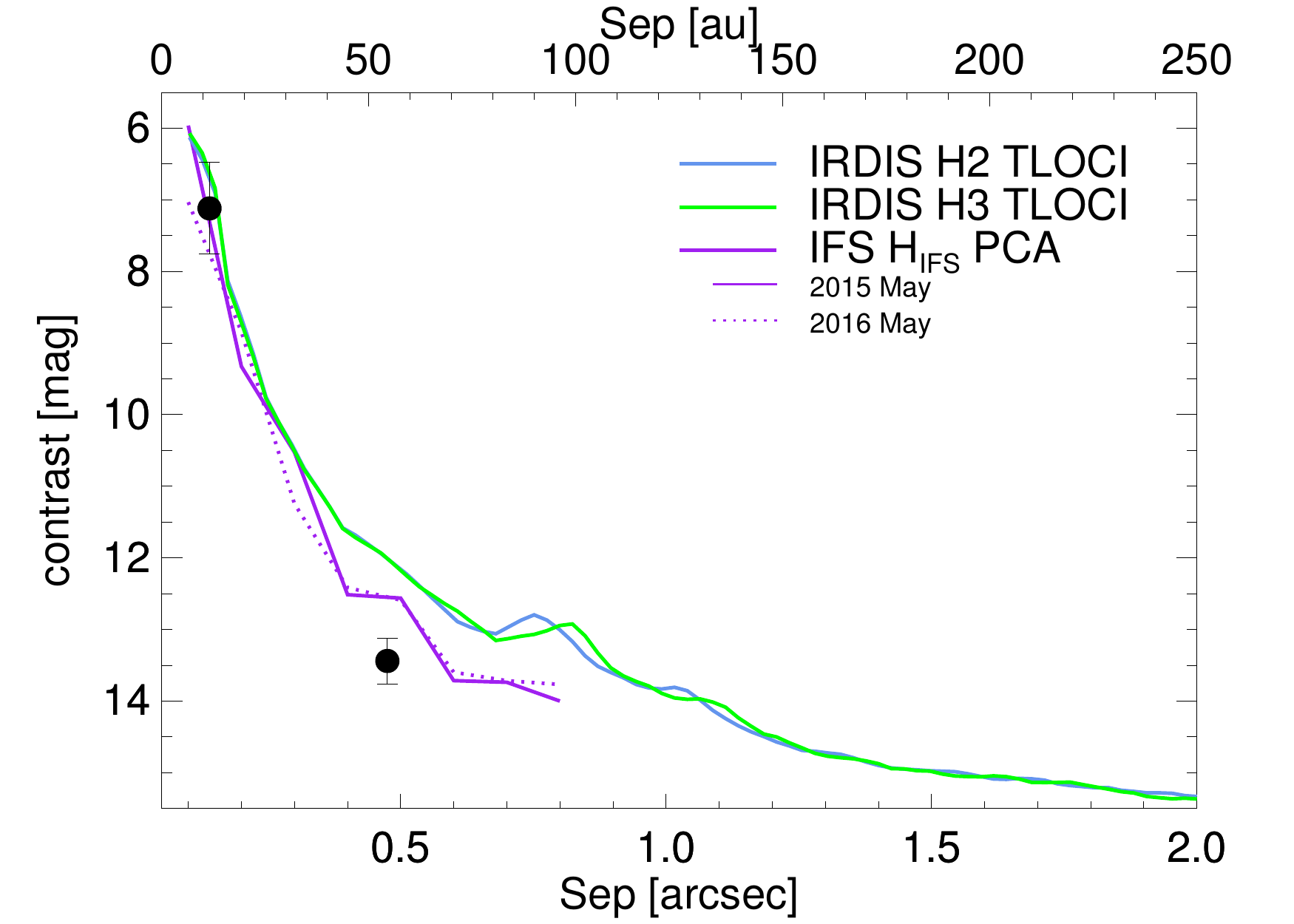} \\
\includegraphics[width=\columnwidth]{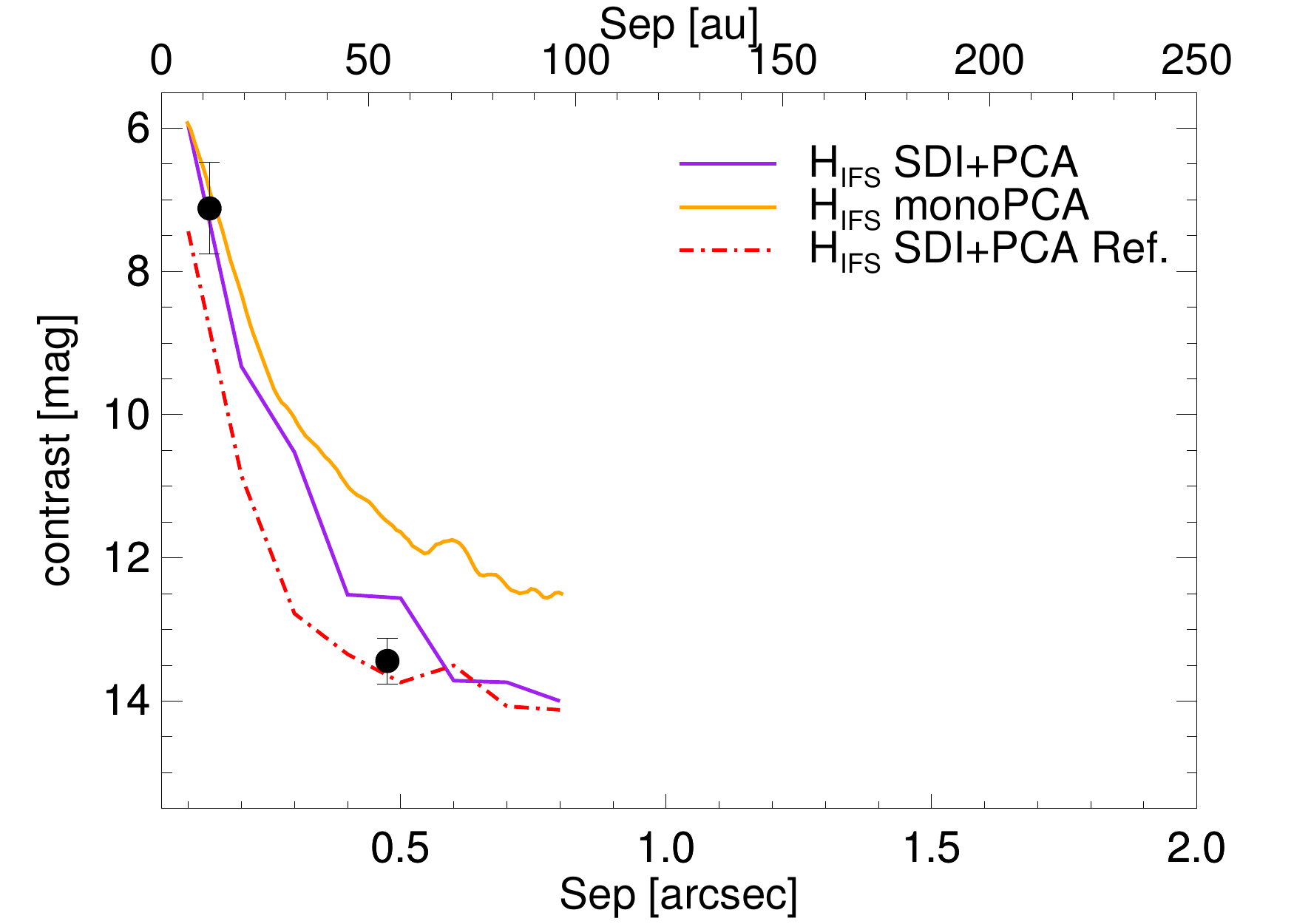}
\caption{Top: contrast curves for HD 100546 obtained in the $H_{IFS}$  band for the two best IFS datasets applying SDI+PCA and for IRDIS H2H3 dataset applying TLOCI. Bottom: The IFS May 2015 contrast curve shown above is compared with the result obtained with 2 modes monochromatic PCA (orange curve) and with the contrast curve in the $H_{IFS}$  band for HD~95086 applying SDI+PCA (in red dashed-dotted line). The planets contrast values measured in the H band obtained by Currie et al. 2015 with GPI are reported in both panels.}
\label{contrast_curves}
\end{figure}

Two candidate companions around HD~100546 have been proposed over the years \citep{2013Quanz1, 2014Brittain, 2015Quanz,2015Currie}. We investigated IFS and IRDIS images to find signatures of the already proposed candidate companions or new signatures. None of the IFS images shows evidence of new candidate companions above threshold while in the IRDIS images seven objects were identified as background stars (see Appendix~\ref{sec:bkg_objects}).

To properly discuss our data, we should consider the detection limits for point sources on our data. The IRDIS $5\sigma$ detection limits for point sources were obtained through the SpeCal software \citep{galicher2018}: it estimates the noise level $\sigma$ as the azimuthal standard deviation of the flux in the reduced image inside FWHM/2 wide annuli at increasing separations, and rescales it as function of the stellar flux derived from the off-axis PSF images, taking into account the transmission of the neutral density filter used to avoid the star image saturation. We evaluated the IFS $5\sigma$ contrast limits in the PCA images following the method described in \cite{mesa2015}. In summary, the algorithm determines for each pixel the standard deviation of the flux in a $1.5\times 1.5\lambda/D$ box, centred on the pixel itself and divides it by the proper stellar flux (as described above). For a given separation, the $5\sigma$ is then computed as five times the standard deviation of the values obtained for all the pixels at that separation. Finally both IRDIS and IFS $5\sigma$ values are corrected for the algorithm throughput and for the small number statistics \citep{mawet2014}. To this purpose, fake companions, ten times more luminous than the noise residuals in the final reduced image, were injected in the pre-processed frames at various separations from the star and the datacube is reduced as before. This is then repeated at several different position angles and the throughput final value is the average of fake planets flux depletion at the same distance, corrected for the coronagraph attenuation.

In Fig. \ref{contrast_curves} (Top) we show the deepest 5$\sigma$ contrast curves obtained with IRDIS and IFS for HD~100546 in the H band. The TLOCI-based analysis shows that IRDIS could reach a $5\sigma$ contrast $>12$ mag at separation $>0.5$\arcsec\ and even 15 mag at separations of 1.5\arcsec. in both H2 and H3 bands  The IFS deepest contrast is reached when applying SDI and PCA. However, all these approaches, widely used for isolated point-like sources, likely underestimate the contrast limits in the presence of a bright disk such as that of HD~100546, because the disk flux enters both in the estimation of the background and in the attenuation effect and therefore a less aggressive approach (e.g. monochromatic PCA with low number of modes) is more suitable. We considered the HD 95086 data set mentioned above to show what will be the expected contrast limit in absence of the disk (red dash-dotted line in Fig.~\ref{contrast_curves} In this Figure, we also show the H magnitudes of the two candidate companions around HD~100546 from \citet{2015Currie} using GPI: the impact of the disk on the contrast limit for IFS images is not negligible at separation of both CCc ($\sim 2.2$ mag) and CCb ($\sim 1.5$ mag). We note here that CCc with a contrast of 7.12 mag should be nominally visible in the IFS May 2015 and May 2016 data sets, where the contrast limits at the candidate separation are 7.3 mag and 7.8 mag, respectively. CCb ($\Delta H=13.44$, $r\sim0.47$\arcsec), instead, is below the detection limit in the H band. However it would be above the detection limit if we consider the limits applicable to a star without a luminous disk.

\subsection{CCc}

\begin{figure}
\centering
\includegraphics[width=\columnwidth]{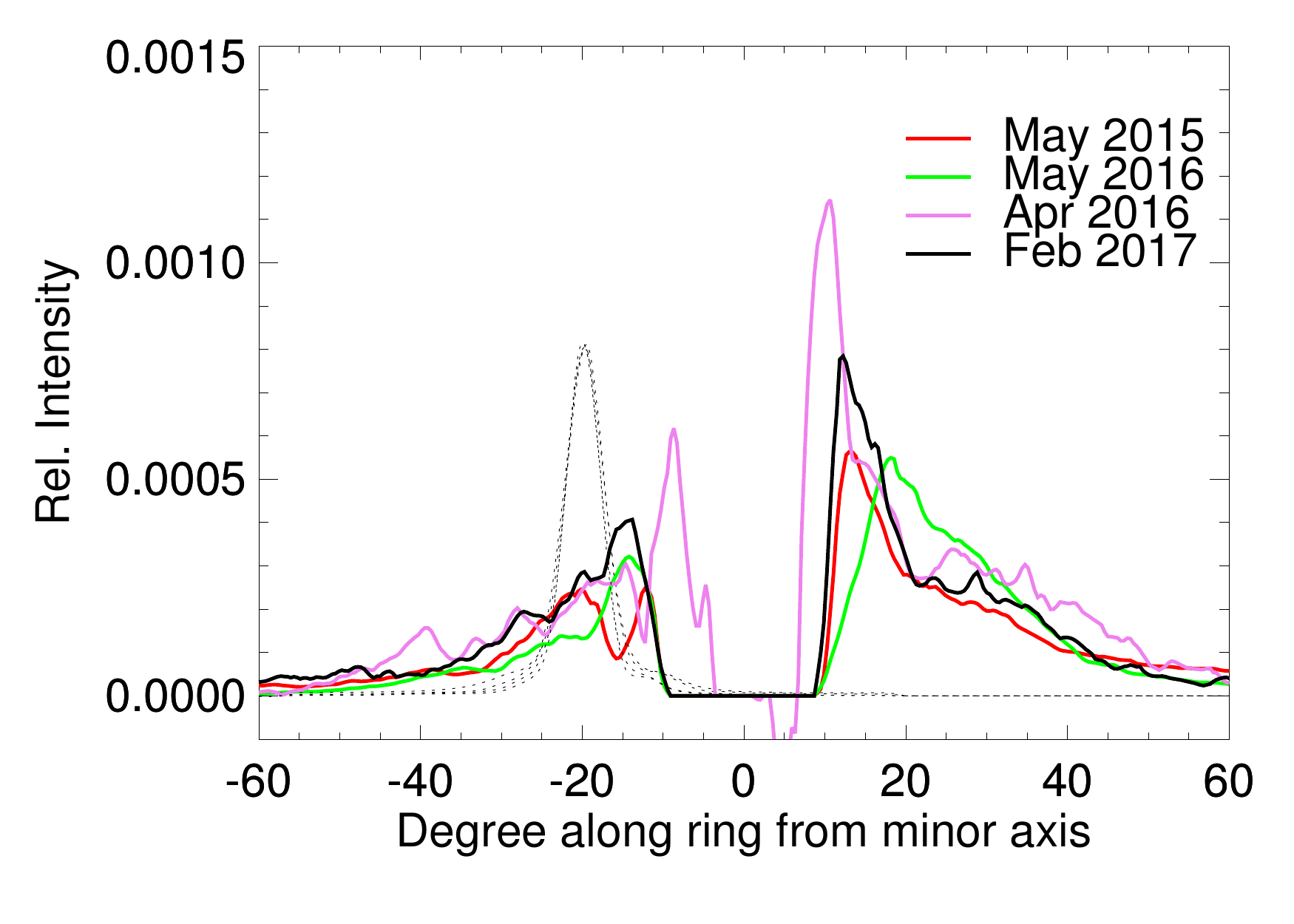} 
\caption{Intensity profile as a function of phase of the two wings at different epochs. Phase 0 correspond to the near side of the semi-minor axis and negative values correspond to the SE wing. The dashed profile correspond to a point like source located at the expected position of CCc crossing the ring with the contrast upper limit given by \cite{2015Currie}. April 2016 data set was scaled by an arbitrary factor of 1.6 to overlap other profiles.}
\label{wing_brigthness}
\end{figure}

No obvious point-like structure is visible at the expected position of CCc in any of our IFS and IRDIS data sets in all epochs. Using a different reduction approach, \cite{2016Garufi} identified a bright knot in the SE arm ($r\sim 0.120 \pm0.016$\arcsec, $PA\sim165\pm6$\degree) at least in H band for the May 2015 data. On one hand, one could try to associate CCc with this knot, as it may lie at/just inside the disk cavity and is counterclockwise from the position noted by \cite{2015Currie}, as expected for an orbiting object predicted from \cite{2013Brittain}. On the other hand, these results do not establish CCc as a companion since there are equally compelling alternatives. We detect bright, elongated emission at a similar location in all our IFS and IRDIS H2H3 data set, suggesting confusion between bright disk emission and that of any point source. Moreover, all our data sets have a light distribution along the two wings that is nearly symmetric around the disk semi-minor axis, with the NW side being more luminous, as already found by \cite{follette2017}. The emission could simply be a non-polarized hot spot in the disk. Furthermore, this structure is detected at locations compatible with a stationary object with respect to the star, despite a planet orbiting the star on the disk plane as detected by \cite{2014Brittain} should move by $\sim15$\degree\ (one resolution element at that separation) during our campaign. The feature is clearly visible only when using not aggressive reduction methods, such as monochromatic PCA with only one component.

A key challenge in interpreting these data is the effect of processing. At CCc’s angular separation, parallactic angle motion is small ($1—2 \lambda/D$ for our data) and self-subtraction due to processing is severe. Processing could anneal the disk wings to look like a point source, (a shock, a convergence of spiral features, etc.). Alternatively, processing might preferentially anneal a point source, making it appear radially elongated and indistinguishable from the disk wings. Additionally, the proximity of CCc to the coronagraph edge when a mask was used might affect both GPI and SPHERE results. With the coronagraph removed, the brightest part of both southern and northern wings shift closer to the minor axis (Fig.~\ref{wing_brigthness}). The SPHERE coronagraph we used, indeed, is 0.03” smaller than that used in GPI H-band observations and this technical difference could justify some different results obtained with the two instruments. Finally, we should consider that the observation taken by Currie et al. (2015) is not simultaneous to IFS and IRDIS observations and the sky conditions were better. This leaves open the additional possibility that we could not recover the same feature detected by Currie et al. (2015) because the orbital motion moved it behind the coronagraph or the circumstellar disk, during this interval of time.

Investigating the nature of the claimed CCc further requires a detailed forward-model of both any point source and the disk over multiple data sets, that takes into account the impact of the coronagraphs, and is beyond the scope of this paper.

\subsection{CCb}
\label{sec:ccb}

\begin{figure}
\includegraphics[trim=0.7cm 7cm 6cm 1.2cm, clip, width=\columnwidth]{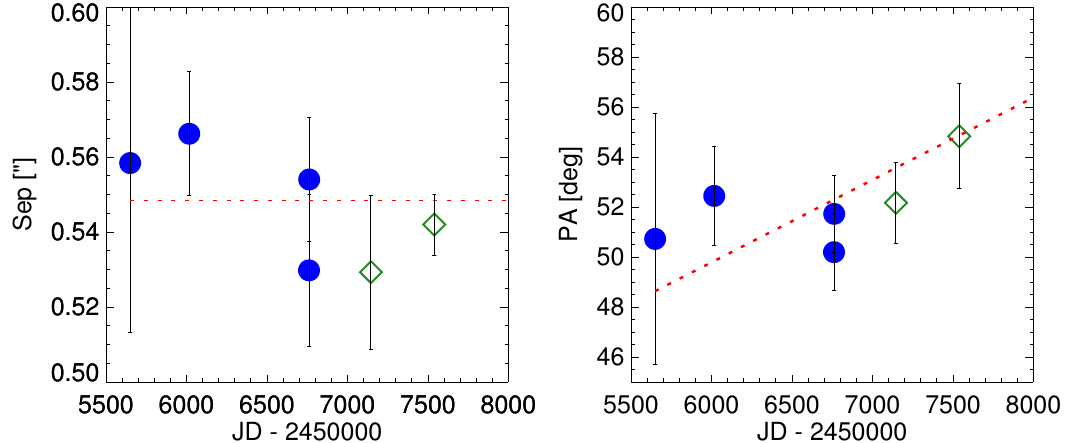}
\caption{Run of deprojected separation (left panel) and of position angle (right panel) of CCb with time. Data are from \cite{2013Quanz1, 2014Currie, 2015Quanz} (blue dots) and this paper (open diamonds). Dashed lines are predictions for a circular orbit on the plane of the disk.}
\label{CCb_Astrometry2} 
\end{figure}

\begin{figure*}
\centering
\includegraphics[width=\textwidth]{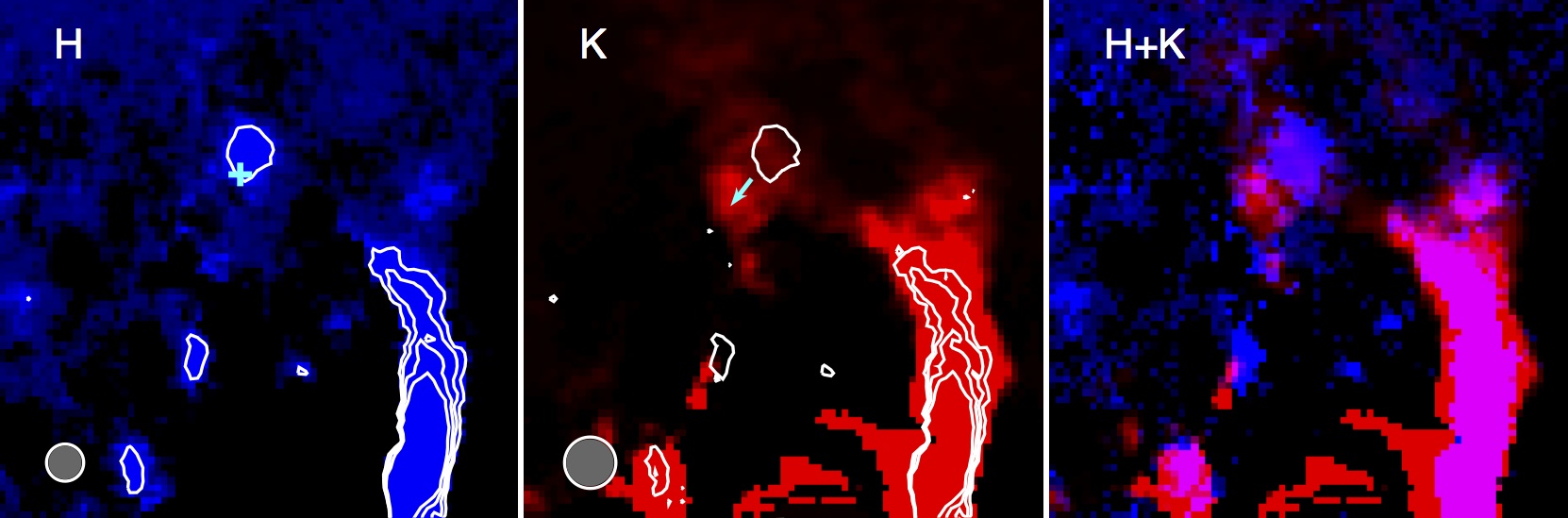}
\caption{Comparison of HD~100546 images in different band filters. One component PCA images of IRDIS H2H3 taken on May 29th 2015 (left), K1K2 taken on May 3rd 2015 (middle) and their combination(right). Contours refer to IRDIS H2H3. The cross represents \cite{2015Currie} detection in GPI H band , the arrow in the central panel indicate the motion of CCb between  \cite{2015Quanz} detection in NACO $L'$ band  and our May 2016 K1K2 detection of CCb. The grey circle represent the resolution element of the images. }
\label{Ccb_coinc_HK} 
\end{figure*}

\begin{table}
\centering
\caption{Astromety of CCb in the SPHERE K band at different epochs.}
\begin{tabular}{lcc}
\hline
Date & $r$[mas] & PA [$\deg$]\\
\hline
\hline
May 3 2015 & $454\pm10$ & $10.4\pm1.5$  \\
May 31 2016 & $456\pm10$ & $12.5\pm1.5$\\
\hline
\end{tabular}
\label{tab:CCbastro}
\end{table}

Since the initial discovery of HD100546 b by \cite{2013Quanz1}, its re-detection has been reported from different groups using different instruments and in multiple wavelengths \citep{2013Quanz1,2014Currie, 2015Quanz}. However, the debate on the nature of CCb was recently reactivated by  the results obtained by \cite{2015Currie} and \cite{rameau2017}, both with GPI H-band observations. In both cases, a clear signal is detected at a location possibly compatible with CCb, but, on one hand \cite{2015Currie, currie2017} identified this with CCb (a  point source over-imposed to a flat disk component), on the other \cite{rameau2017} interpreted it as stellar scattered light, being point-like or extended depending on the processing method used.

In the following discussion we use a not aggressive monochromatic PCA approach with only one component over the whole IFS FoV simultaneously, and over the inner circle of 0.8\arcsec\ for IRDIS one. Moreover, we exploit the very wide spectral range simultaneously offered by the SPHERE IRDIFS\_EXT set up (from 0.95 up to 2.2\,\micron). All IRDIS K1K2 datas ets show a clear diffuse emission on top of the Northern wing (as already noticed by \citealt{2016Garufi} for the 2015 data sets) that is compatible with the CCb detections in $L'$ and $M$ band using NACO. 

In order to obtain a more precise characterization of this feature, we focused on the two best IRDIFS\_EXT data sets and we found out that the position of this source varies a little between the two epochs, as shown in Table \ref{tab:CCbastro}. From  the disk velocity retrieved from spectroastrometric analysis of different molecules \citep[see e.g., ][]{acke2006,panic2010, brittain2009} and/or with ALMA observations \citep{walsh2014},  it was derived that the disk rotates counter-clock wise, with the SW part being the closest to the observers. We discover that the feature we detect is moving counter-clockwise and its astrometry, combined with previous detections, is compatible with a Keplerian motion on the disk plane. This is shown in Fig. \ref{CCb_Astrometry2}, where we present the run of the separation and position angle of this source with time, using data from \citet{2013Quanz1,2015Quanz,2014Currie} and the two best K1K2 epochs from SPHERE. We adopted the disk plane inclination of 42\degree\ and  position angle of 146\degree\ \citep{2014Pineda}, obtaining an orbital radius of $545\pm15$\,mas that corresponds to $59.5\pm2$\,AU at the distance of HD~100546. Given  the stellar mass of 2.4 M$_\odot$ \citep[see e.g. ][]{2013Quanz1}, the corresponding period is 299 yr. We over-imposed the predictions for a similar orbit on the run of separation and position angles shown in Fig. \ref{CCb_Astrometry2}, dashed lines. The agreement  between expectation and observations is fairly good, in view of the rather large error bars associated with all these data. We conclude that the motion of CCb is compatible with a circular orbit on the plane of the disk, in the same direction as the disk rotation, although there is room for different orbital solutions and also for stationariness, given the uncertainties on the positions. If this orbital solution is confidently assessed, we could therefore predict that in Spring 2019 CCb will have moved enough to disentangle the motion from being a stationary object, or it will disappear behind a disk structure. For example, assuming a nominal co-planar, circular orbit at 59.5 au, CCb should appear at separation of $\sim440$ mas and PA $\sim 17$\degree\ in May 2019, which is roughly three resolution elements away from its position in May 2015. However, the astrometric points hint about a slower motion that can be due to an eccentric and/or no-coplanar orbit. If so, the time baseline needed to confirm the CCb movement stretches out.

This object has a contrast of $12.08\pm0.49$\,mag in K1 and $11.68\pm0.60$\,mag in K2, compatible with the NACO non detection by \cite{boccaletti2013}, while its H band upper limit is $13.75\pm0.05$\,mag. On the other hand, the IRDIS H2H3 data set and the H part of the IFS images (simultaneous to the K1K2 IRDIS data sets) show a diffuse emission, located $\sim 70$\,mas NW with respect to the previous one in continuation with the Northern wing, as shown in Fig. \ref{Ccb_coinc_HK}. This emission, with median position  $r=477\pm 12$\,mas, $PA=7.2\pm1.5$\degree, is consistent with the detection in the H-band by \cite{2015Currie} and \cite{rameau2017}. This source is clearly distinct from that detected at longer wavelengths, and looks extended. However, given the angular resolution of GPI and the different H-band filter used in GPI and SPHERE, the detection of \cite{2015Currie, currie2017}, intermediate between the H and K detection with SPHERE, can be interpreted as a combination between these two. Finally, only very weak diffuse emission was visible at wavelengths shorter than 1.1\,\micron.

Additional information on CCb cannot be retrieved by the polarimetric data: no emission is seen in the IRDIS Q$_\phi$ images as shown in Figure \ref{PDI}(h), while in the PDI+ADI image the residuals due of the telescope spider are not negligible and fall at the location of CCb, so it is impossible to tell whether or not a disk structure at that location could be interpreted as a point source after filtering by ADI.

\subsection{Interpreting CCb as extended source}

Previous works suggested that CCb is surrounded by circumplanetary disk \citep{2015Quanz,2015Currie,rameau2017}. In \cite{2015Quanz} the presence of a spatially unresolved circumplanetary disk ($r\sim1.4$ au around a $2 M_{J}$ object) is considered as an explanation of the discrepancy between the observed values of radius and effective temperature with those obtained by models for very young gas giant planets. 

\begin{figure}
\centering
\includegraphics[width=\columnwidth]{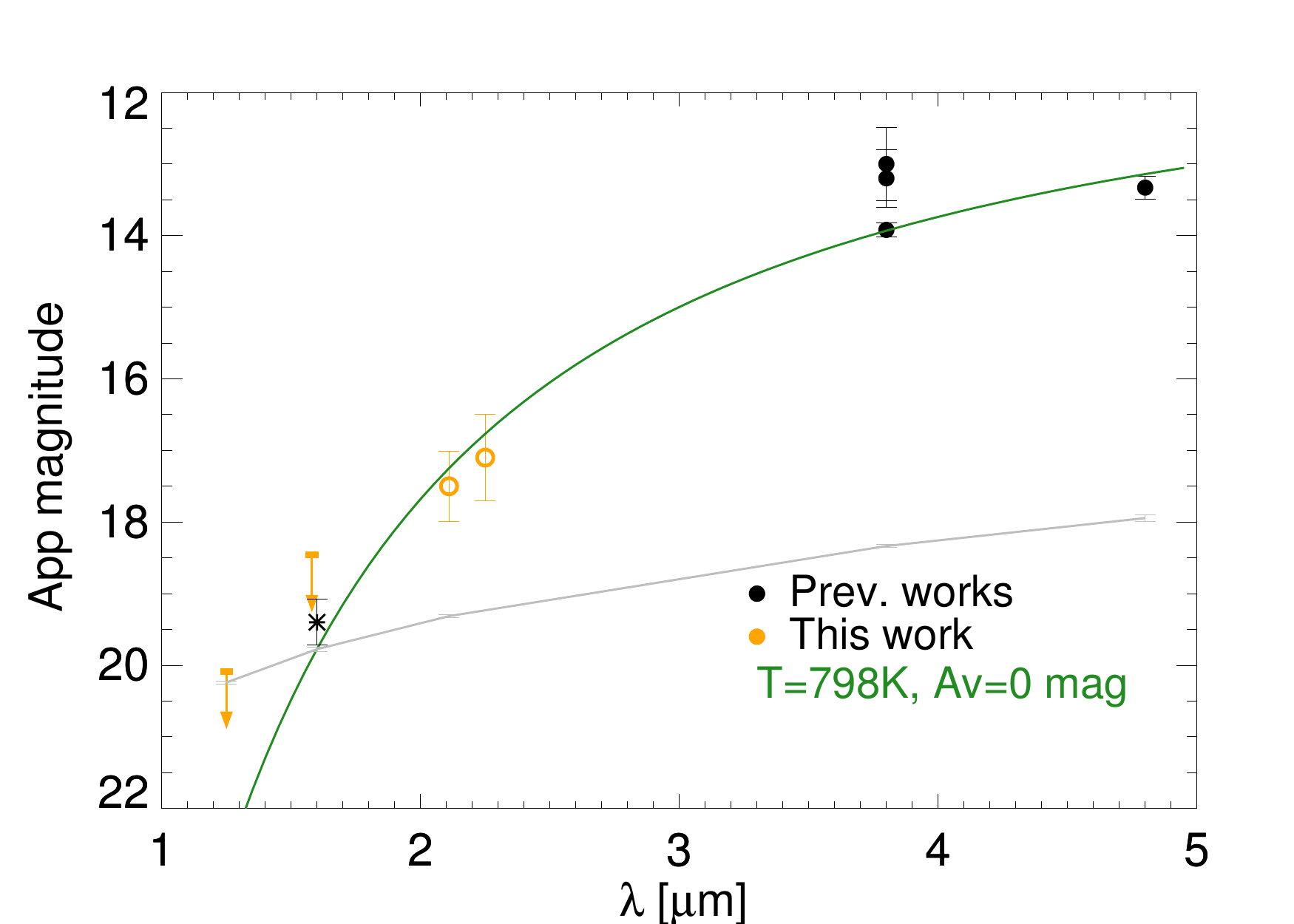} 
\caption{SED of CCb planet and disk combining previous results (black) with our upper limits (orange). The green line represent a black body of 932 K with an absorption $A_V=28$ mag (see text). The grey curve represent a grey contrast of 14 mag fitting the short wavelengths observations.  In particular, only data corresponding to black filled circles were considered for the relative astrometry. The black star symbol refers to apparent magnitude of a source located at about 28 mas ($\sim 0.7 \lambda/D$) from the position of the source detected at longer wavelengths, as given by \cite{2015Currie}.}
\label{magb}
\end{figure}

\begin{table}
\begin{minipage}{\columnwidth}
\centering
\caption{At different wavelengths, we report the star magnitude $m_*$, the contrast of CCb or it upper limit ($cont_b$) and the associated uncertainty ($err_{cont}$). The values obtained by this work are not corrected for any dust extinction.}
\begin{tabular}{lrrcc}
\hline
$\lambda$ &  $m_*$ & $cont_b$ &$m_b$  &  Reference\\
$\mathrm{[\mu m]}$ & $\mathrm{[mag]}$&$\mathrm{[mag]}$&$\mathrm{[mag]}$& \\
\hline
\hline
4.8  & 4.13  & 9.2  & 13.33$\pm$0.16   & 1 \\
3.8  & 4.2 & 9.0 & 13.2 $\pm$0.4   & 2 \\
3.8  & 4.52  & 9.4 & 13.92$\pm$0.1    & 1 \\
3.8  & & &13.06$\pm$0.51    & 3 \\
\hline
2.25 & 5.42 & 11.68 & 17.10 $\pm$0.60 & 4\\
2.2  & 5.42  & $>9.09$  &  $>14.51$   & 5 \\
2.1  & 5.42  & $>9.60$   & $>15.02$   &  1 \\
2.11 & 5.42 & 12.08 & 17.50 $\pm$0.49 &4\\
\hline
1.60 &  & & 19.40 $\pm$ 0.32$^a$  & 6\\
1.58  & 5.96  & $>12.5$ & $>18.46$ &    4 \\
\hline
1.25  &  6.42 & $>13.6$ & $>20.02$ &  4  \\  
\hline
\end{tabular}
\label{tab:magb}
\end{minipage}
\tablefoot{1 \cite{2015Quanz}, 2 \cite{2013Quanz1}, 3 \cite{2014Currie}, 4 this work, 5 \cite{boccaletti2013}, 6 \cite{2015Currie}.\\ $^a$This value refers to the apparent magnitude of the source identifyed by \citep{2015Currie} and located at about 28 mas ($\sim 0.7 \lambda/D$) from the expected position of the source detected at longer wavelengths.} 
\end{table}

At wavelengths shorter than 2\,\micron, we evaluated the 5$\sigma$ limit to the magnitudes of a point source at the presumed location of CCb. With IFS we obtain that it is fainter than the apparent magnitude 19.84 in the J band ($1.20-1.30$\,\micron), fainter than 19.80 mag in J broad band ($1.15-1.35$\,\micron) and fainter than 19.05 mag in H; uncertainties on these values came from the noise distribution estimation at the CCb location.

Putting together all the results and the literature ones (see Tab. \ref{tab:magb}) we obtain the CCb SED plotted in Fig.~\ref{magb}. Some important arguments that one should take into account interpreting this SED are: (a) CCb is below detection threshold in our IFS images, so we only estimated the contrast limits in the corresponding area of CCb, while in the H2H3 filter we only detect a very weak diffuse emission at its location; (b) what we see in the IRDIS data at positions corresponding to CCb could be simply a bright structure of the disk; (c) we are  not subtracting the disk contribution to the flux. Similar to what discussed above, the self subtraction on extended sources is not negligible even with this not aggressive reduction method and it is difficult to estimate because it depends on the specific distribution of light than cannot be easily modelled. This implies that the ADI technique alters the photometry of the extended object. What we can conclude is simply that the structure seen in our K1K2 images is compatible with an extended object whose apparent magnitude is brighter than what is expected for a point source at the same location and is compatible, within the uncertainties, with the flux observed at longer wavelengths.

 Given the small difference of contrast between the J/H-band diffuse source and the strong difference with the L/M-band putative point-source, we may represent this emission as the sum of two different sources: a very red compact source and a more extended one with a flatter spectrum. It is clear in fact that this source is redder than the star and cannot be explained as stellar scattered light alone, since a flat contrast (grey line) that fits the short wavelength points largely fails to reproduce the L and M data. The green line represents the SED of a black body with $T_{eff}\sim800$\,K and no absorption, that implies a radius of the emitting area $R=12.5\,R_J$. Taking into account the new determination of HD~100546 distance, this result is in fair agreement with  \citealt{2015Quanz} ($T_{eff}=932^{+193}_{-202}$\,K, $R=6.9^{+2.7}_{-2.9}\,R_J$). It is noteworthy that, given the small number of photometric points and limited spectral range, absorption and temperature are degenerate, and therefore many combination of temperature and reddening can well reproduce the SED. However, given its very young age, the object likely does not emit as a black-body, but the accreting disk shock contribution to the total luminosity is not negligible, and could also be dominant \citep{mordasini2017}. Also the filter used are built to provide information on absorbing molecules and therefore can play a role in the flux estimation at given wavelengths.

The luminosity of a forming and accreting planet in different formation scenarios is evaluated in \cite{mordasini2017}. They analyzed the case of HD~100546 b in detail, starting from the physical parameters derived by \cite{2015Quanz}, and could constrain the mass of this source only with large uncertainties, due to different mechanisms allowed. To disentangle the emitting sources and hence estimate the CCb mass, we need to better characterize the SED, that can be possible with space observations with new facilities (like JWST) or with the ELT class instruments. Alternatively, a better spatial resolution will allow to resolve the circumplanetary region and retrieve the mass of CCb through dynamical models.

The nature of the extended emission detected in the H2H3 filter is not clear. It can be the either a source physically bound to CCb, or a bright structure of the disk. Further deep observations  and/or a longer (3-4 yr) temporal coverage can disentangle between these possibilities. If we assume that they are an unique extended source, we obtain that the radius is of the order of $r$=$34$ mas ($\simeq3.7$ au), compatible with previous estimations. If this were a circumplanetary disk, the corresponding Hill's radius ($R_{Hill}$) of the planet would be expected to be about 3 times larger, following models by e.g. \citet{shabram2013, ayliffe2009, quillen1998}. Therefore, the observed structure is compatible with a circumplanetary material around a massive planet that could be responsible for carving the gap as demonstrated by hydrodynamical models \citep[e.g.][]{2015Pinilla, dong2016}.

To justify the extended emission, we further consider a spherical cloud around CCb, located at a separation $d$=$65\pm5$ au from the star, that reflects stellar light. If its optical depth is $\tau>>1$, then we expect that the total reflected light is $c$=$A\pi r^2/(4\pi d^2)$ where $A$ is the albedo. If $A$=$0.5$ then the expected contrast is $c$=$4.2\times10^{-4}$, that corresponds to 8.6 mag, in agreement with our observations. This feature would be unresolved at L and M wavelength and therefore contributes to the flux of the compact source but it should be far less than the contribution due to the companion: its luminosity is compatible within the uncertainties with reflected stellar light from the J to the K band.

We finally notice that, in this scenario we are not considering the absorption due to the circumstellar disk material between the star and the circumplanetary disk. This depends on the thickness of the circumstellar disk and on the height of the planet over the circumstellar disk plane at the epoch of our observations. This effect, and the presence of shadows due to the circumstellar disk that may affect the amount of light incident on a circumplanetary disk, could be not negligible and possibly cause an irregular illumination of the circumplanetary disk. Of course the truth can be a combination of all these factors.

We conclude that the origin of this emitting area is still unclear. While its appearance and the SED are compatible with a highly reddened substellar object surrounded by a dust cloud, we cannot exclude other interpretations, such as the super imposition of two spiral arms, the northern wing and the small IRDIS North arm (see Fig.~\ref{IFSIRDIS_simpleADI}), or disk material flowing to a planet due to its perturbation induced on the disk.

\section{Conclusion} 
\label{sec:concl}

We observed HD~100546 with SPHERE using its subsystems IRDIS and IFS in direct imaging and in polarimetry. Our observations confirm the presence of a very structured disk and reveal additional features. The different post processing techniques reveal different characteristics of this complicated disk. RDI and PDI images are dominated by the almost symmetric intermediate and outer disks, while more aggressive differential imaging technique tell a different story, featuring strongly de-centred rings and spirals. These two views can be reconciled in a picture as follows: 
\begin{itemize}
\item The two bright wings, dominant structures in the IR at separations closer than 500 mas, are a unique structure. This is quite evident from the non coronagraphic images. The presence of the coronagraph and the use of the ADI technique contributed to cancel out the light in the rings region closest to the star. 
\item The new PDI data confirms the presence of a unique arm warping for 540\degree. In the innermost regions, three small spiral arms are detected in both J and K band.
\item Modelling a geometrical representation of the disk coupled with an analytic scattering function, we obtain that the disk rings 1 and 2  extend between $\sim15$~au and $\sim40$~au and between $\sim110$~au and $250$~au respectively, consistent with previous results. The inner edge of the ring 2 and the outer edge of the ring 1 correspond to resonances 3:2 and 1:2 with a 70~au radius orbit, suggesting the presence of a massive object located at that separation.
\item  We do not exclude the presence of additional spiral arms inside the disk rings. In particular, we confirm detection of the two possible spiral arms East and South of HD~100546 previously identified by \cite{follette2017}.
\item The spectrum of this disk does not show obvious evidence for segregation of dust of different size and is well explained by micron sized particles.
\end{itemize}
For what concerns the planets, we have no clear evidences of the CCc detected by \cite{2015Currie}; processing could both cause the disk wings to look like a point source or anneal a point source to be indistinguishable from disk emission. This does not exclude that the planet has moved behind the disk in the time between \cite{2015Currie} observations and the time SPHERE data were acquired.

We identify a spatially diffuse source in K band broadly consistent with CCb. When combined with previous measurements, its photometry is consistent with a blackbody-like emitting source of $\sim800$ K, compatible with a highly reddened massive planet or brown dwarf surrounded by a dust cloud or its circumplanetary disk. Its astrometry might have revealed evidence for orbital motion, a result that can be confirmed with future observations. This object can indeed be the disk perturber suggested by the disk modelling. However, other hypothesis are also possible, such as the overimposition of two spiral arms at the location of the L' and M' detections. 

\begin{acknowledgements}
The authors thank the anonymous referee for a very constructive
referee report that improved the initial manuscript.
The authors thank the ESO Paranal Staff for support for conducting the observations. The authors thank Sascha Quanz, Adriana Pohl and Tomas Stolker for the very useful comments that improved a lot the quality of the paper. E.S., R.G., D.M., S.D. and R.U.C. acknowledge support from the "Progetti Premiali" funding scheme of the Italian Ministry of Education, University, and Research. E.R. is supported by the European Union's Horizon 2020 research and innovation programme under the Marie Sk\l odowska-Curie grant agreement No 664931. This work has been supported by the project PRIN-INAF 2016 The Cradle of Life - GENESIS-SKA (General Conditions in Early Planetary Systems for the rise of life with SKA). The authors acknowledge financial support from the Programme National de Plan\'{e}tologie (PNP) and the Programme National de Physique  Stellaire (PNPS) of CNRS-INSU. This work has also been supported by a grant from the French Labex OSUG@2020 (Investissements d'avenir - ANR10 LABX56). The project is supported by CNRS, by the Agence Nationale de la Recherche (ANR-14-CE33-0018). This work is partly based on data products produced at the SPHERE Data Centre hosted at OSUG/IPAG, Grenoble. We thank P. Delorme and E. Lagadec (SPHERE Data Centre) for their efficient help during the data reduction process. SPHERE is an instrument designed and built by a consortium consisting of IPAG (Grenoble, France), MPIA  (Heidelberg, Germany), LAM (Marseille, France), LESIA (Paris, France), Laboratoire  Lagrange (Nice, France), INAF Osservatorio Astronomico di Padova (Italy),  Observatoire de Gen\`{e}ve (Switzerland), ETH Zurich (Switzerland), NOVA (Netherlands), ONERA (France) and ASTRON (Netherlands) in collaboration with ESO. SPHERE was funded by ESO, with additional contributions from CNRS (France), MPIA (Germany), INAF (Italy), FINES (Switzerland) and NOVA (Netherlands). SPHERE also received funding from the European Commission Sixth and Seventh Framework Programmes as part of the Optical Infrared Coordination Network for Astronomy (OPTICON) under grant number RII3-Ct-2004-001566 for FP6 (2004-2008), grant number 226604 for FP7 (2009-2012) and grant number 312430 for FP7 (2013-2016).
\end{acknowledgements}



\appendix
\section{Background Objects in the IRDIS field of view}
\label{sec:bkg_objects}
\begin{figure*}
\centering
\hspace{-1cm}
\begin{tabular}{cccc}
\includegraphics[trim={0.cm 0.5cm 1cm 0cm},clip,width=0.23\textwidth]{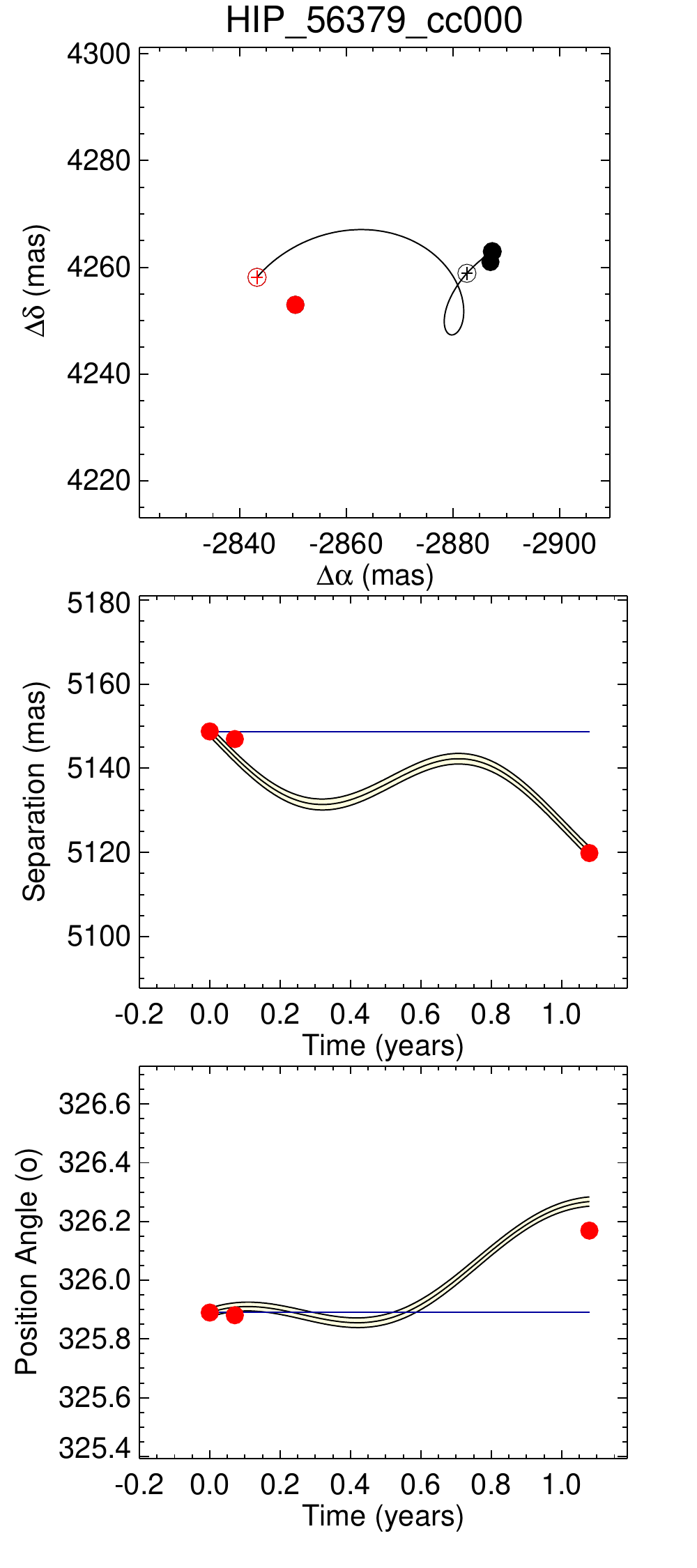} & 
\includegraphics[trim={0.cm 0.5cm 1cm 0cm},clip,width=0.23\textwidth]{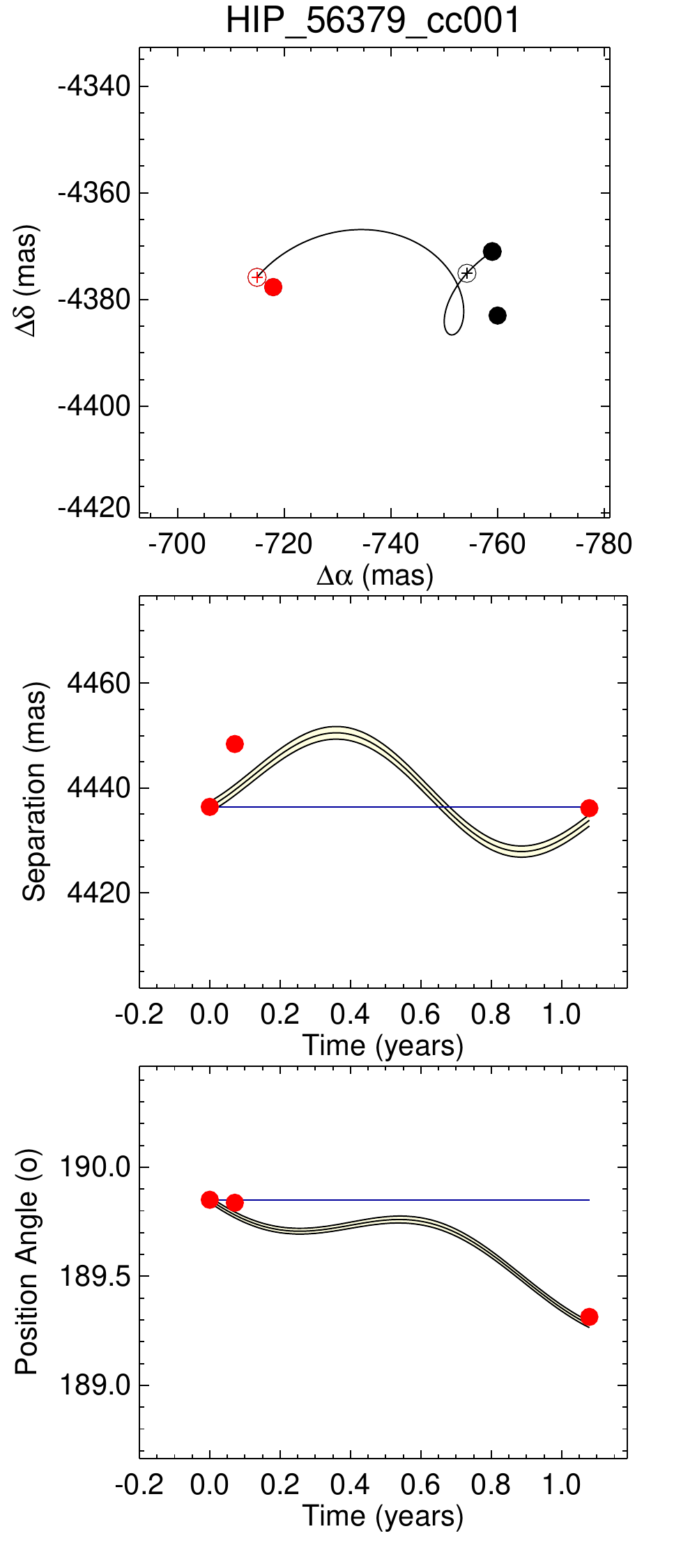}&
\includegraphics[trim={0.cm 0.5cm 1cm 0cm},clip,width=0.23\textwidth]{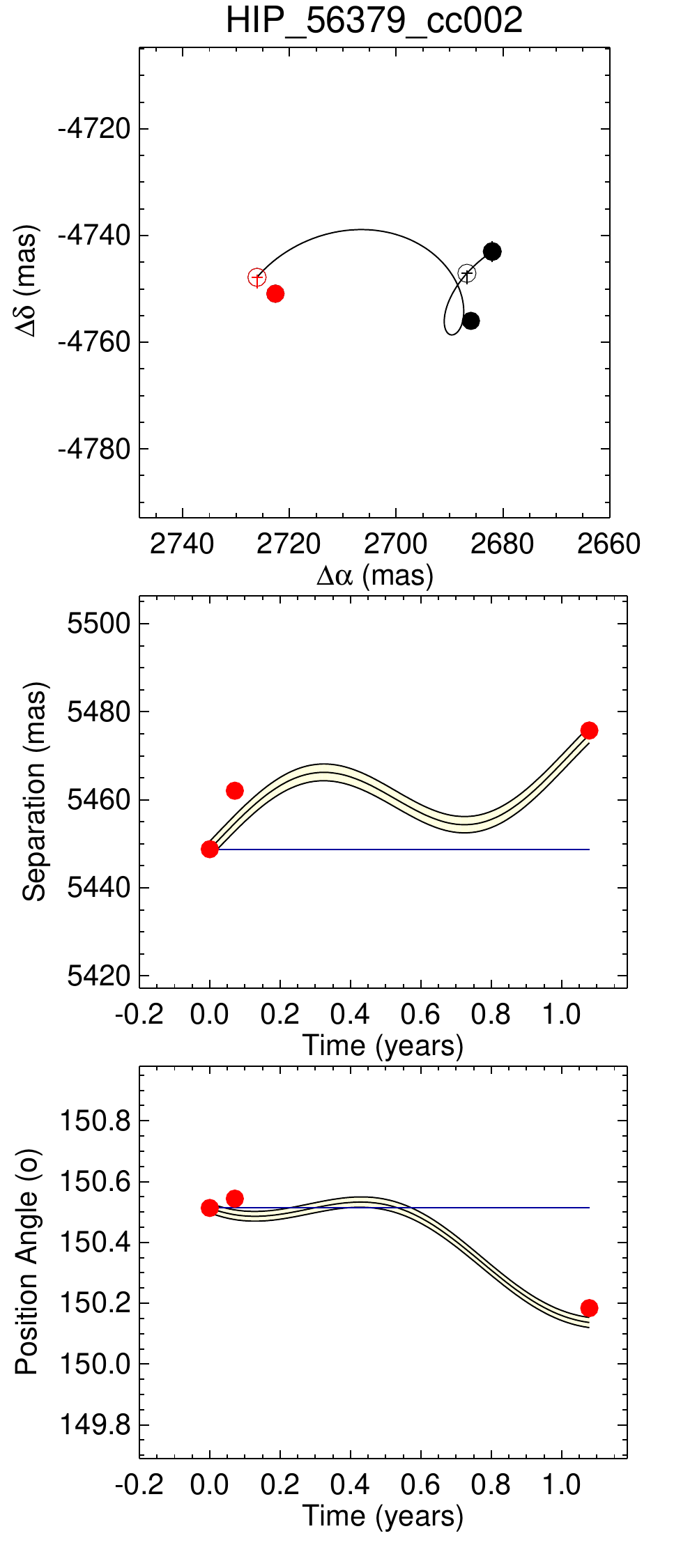}&
\includegraphics[trim={0.cm 0.5cm 1cm 0cm},clip,width=0.23\textwidth]{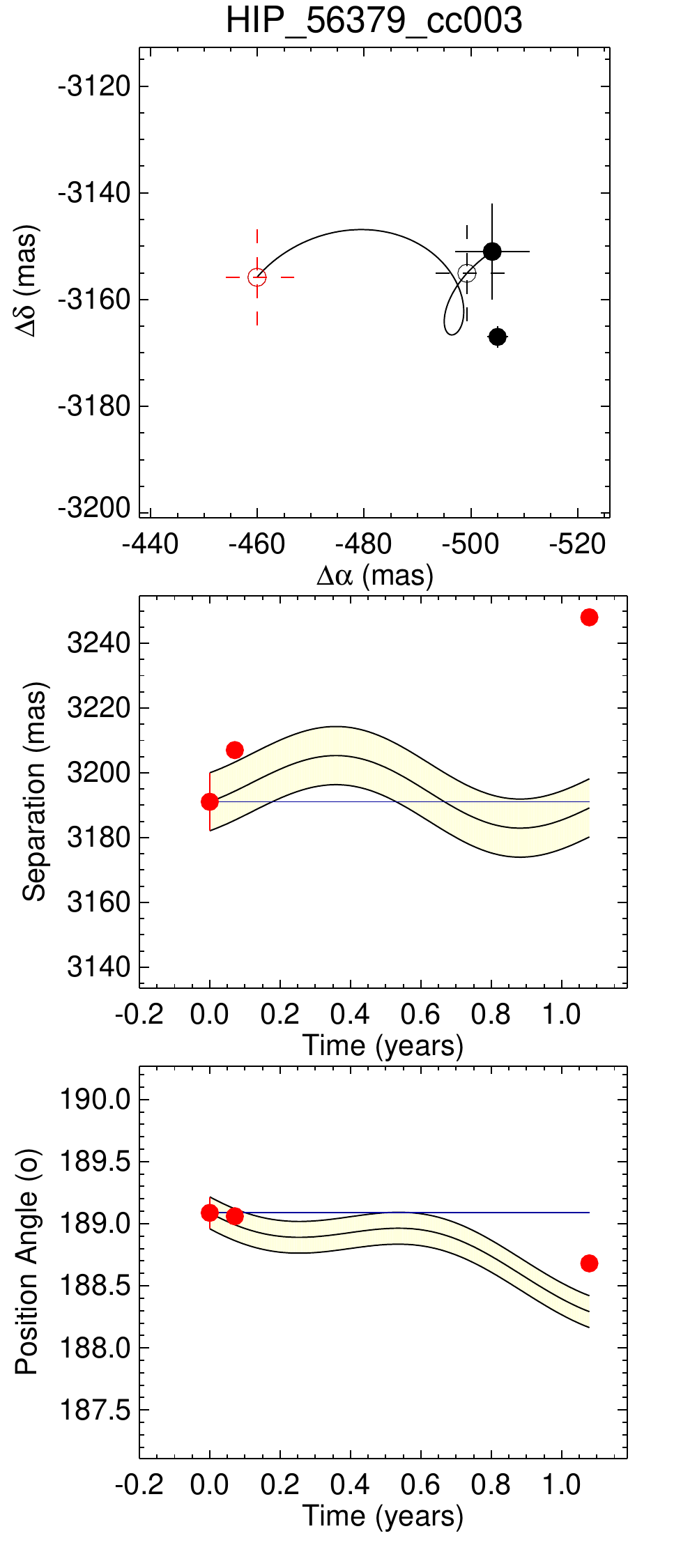}\\
\includegraphics[trim={0.cm 0.5cm 1cm 0cm},clip,width=0.23\textwidth]{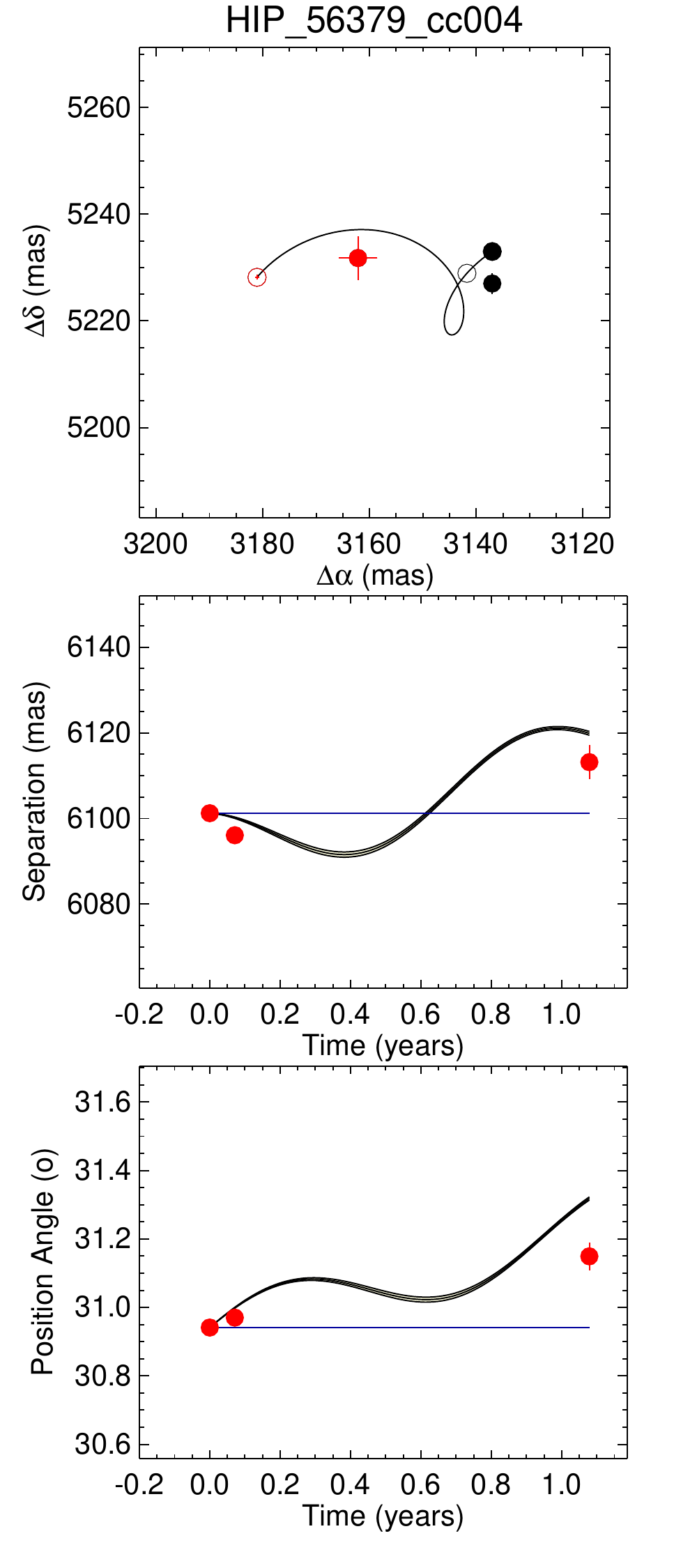}&
\includegraphics[trim={0.cm 0.5cm 1cm 0cm},clip,width=0.23\textwidth]{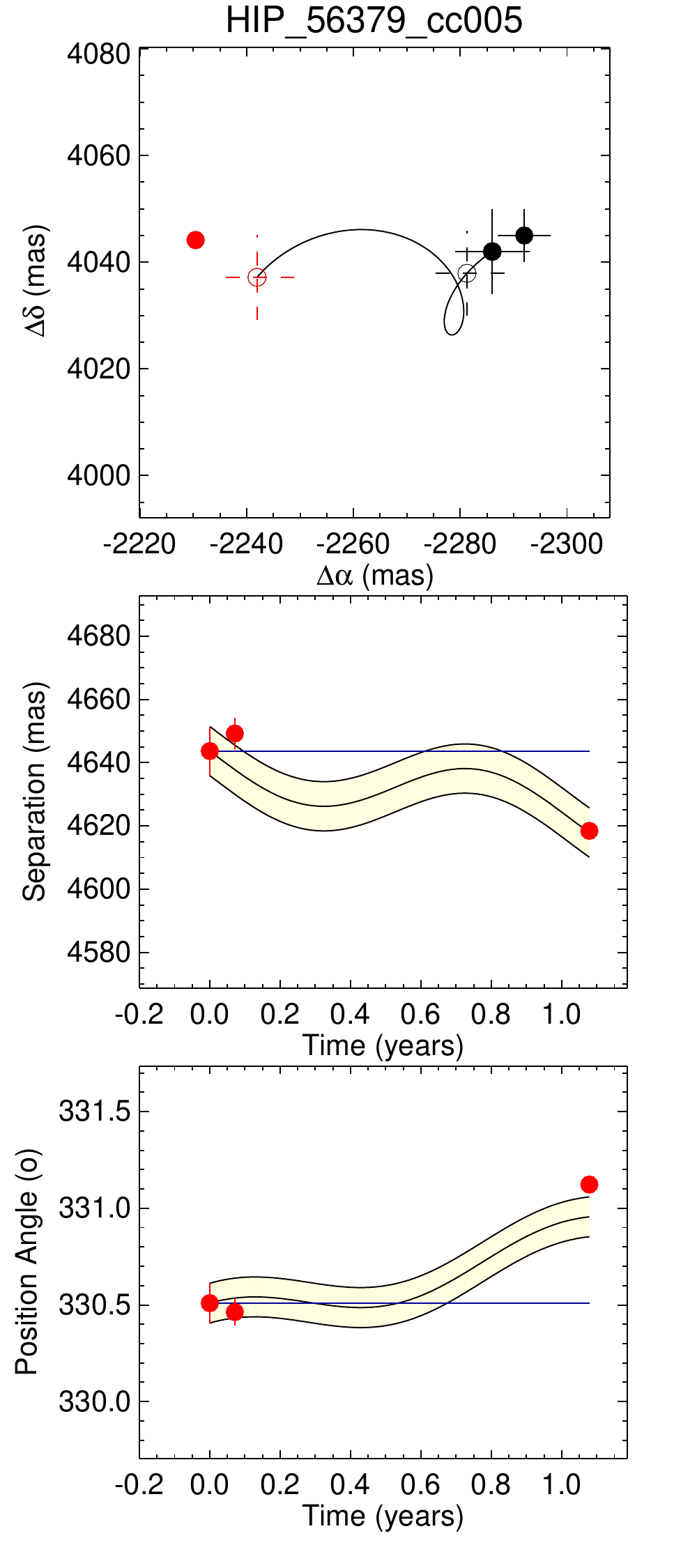}&
\includegraphics[trim={0.cm 0.5cm 1cm 0cm},clip,width=0.23\textwidth]{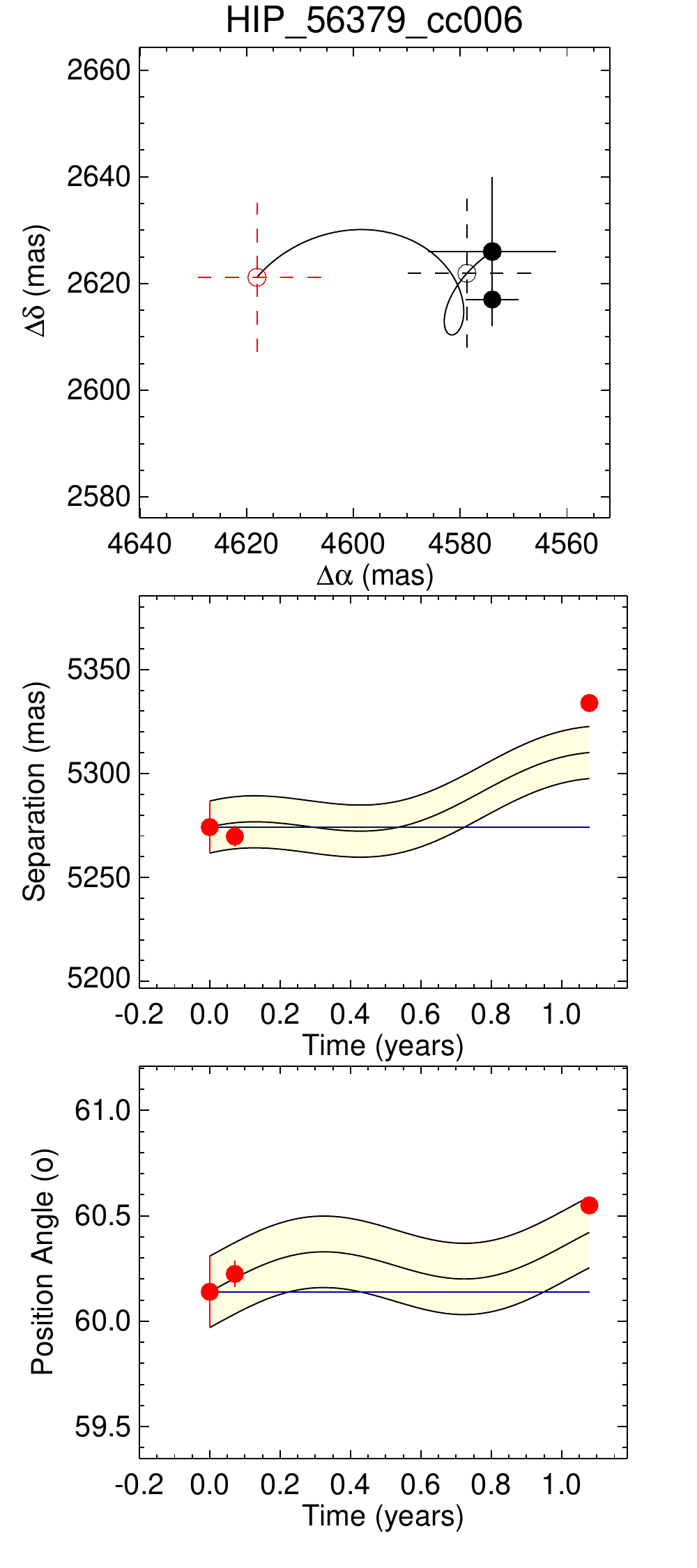}&\\
\end{tabular}
\caption{Top panel: proper motion in $\alpha$ and $\delta$ of the background objects in the IRDIS FoV of HD100546 considering data from 4 May 2015 (filled black), May 29th 2015 (open black), May 31st 2016 (filled red). Dotted points and errorbars represent the expected position for a background object. Middle panel: observed time variation of the separation compared with the expected one for a background object. Bottom panel: observed time variation of the position angle compared with the expected one for a background object. }
\label{CC7}
\end{figure*}
\begin{table}
  \centering
  \caption{Separation in $\alpha$ and $\delta$ of the background objects in the IRDIS FoV of HD 100546 considering data from 4 May 2015, May 29th 2015 and May 31st 2016. Epochs with Italic shape values correspond to those observed points not represented in Figure \ref{CC7}.}
    \begin{tabular}{rrrrrr}
    \hline
    date  & cc\_id & $\Delta RA$   & $\Delta RA_{err}$ & $\Delta \delta$  & $\Delta \delta_{err}$ \\
    & & [mas]& [mas]& [mas]& [mas]\\
    \hline
    \hline
    5/4/15 & 0     & -2887 & 1     & 4263  & 1 \\
    5/30/15 & 0     & -2887 & 1     & 4261  & 1 \\
    6/1/16 & 0     & -2850 & 1     & 4253  & 2 \\
    \hline
    5/4/15 & 1     & -759  & 1     & -4371 & 1 \\
    5/30/15 & 1     & -760  & 1     & -4383 & 1 \\
    6/1/16 & 1     & -718  & 1     & -4378 & 1 \\
    \hline
    5/4/15 & 2     & 2682  & 1     & -4743 & 2 \\
    5/30/15 & 2     & 2686  & 1     & -4756 & 1 \\
    6/1/16 & 2     & 2723  & 1     & -4751 & 1 \\
    \hline
    5/4/15 & 3     & -504  & 7     & -3151 & 9 \\
    5/30/15 & 3     & -505  & 2     & -3167 & 2 \\
    6/1/16 & 3     & \it{-490}  & 0     & \it{-3211} & 0 \\
    \hline
    5/4/15 & 4     & 3137  & 0     & 5233  & 0 \\
    5/30/15 & 4     & 3137  & 1     & 5227  & 2 \\
    6/1/16 & 4     & 3162  & 4     & 5232  & 4 \\
    \hline
    5/4/15 & 5     & -2286 & 7     & 4042  & 8 \\
    5/30/15 & 5     & -2292 & 5     & 4045  & 5 \\
    6/1/16 & 5     & -2230 & 0     & 4044  & 0 \\
    \hline
    5/4/15 & 6     & 4574  & 12    & 2626  & 14 \\
    5/30/15 & 6     & 4574  & 5     & 2617  & 5 \\
    6/1/16 & 6     & \it{4645}  & 0     & \it{2623}  & 0 \\
    \hline
\end{tabular}
\label{tab:addlabel}
\end{table}
As described in Sect.~\ref{IFSIRDIS_simpleADI}, seven points sources are detected in the IRDIS FoV. Their astrometry and photometry are listed in Table~\ref{CC7} and  Fig.~\ref{CC7} unambiguously shows that all of them are background objects. 

\label{lastpage}
\end{document}